\documentclass[twocolumn,preprintnumbers,amssymb,amsmath,9pt]{revtex4}[10pt]
\usepackage{graphicx}
\usepackage{dcolumn}
\usepackage{bm}
\usepackage{amssymb}
\usepackage{epsfig}
\newcommand{\be}{\begin{equation}}
\newcommand{\ee}{\end{equation}}
\newcommand\beq{\begin{eqnarray}}
\newcommand\eeq{\end{eqnarray}}

\newcommand{\Cov}{{\rm Cov}}
\newcommand{\fsky}{f_{\rm sky}}
\newcommand{\beam}{B_\ell}

\begin{document}

\title{Constraining Models of Neutrino Mass and Neutrino Interactions with the Planck Satellite}

\preprint{LA-UR-07-2135, MADPH-07-1483}

\author{Alexander Friedland}
\affiliation{Theoretical Division, T-8, MS B285, Los Alamos National Laboratory, Los Alamos, NM 87545}
\author{Kathryn M. Zurek}
\affiliation{Department of Physics, University of Wisconsin, Madison, WI 53706}
\author{Sergei Bashinsky}
\affiliation{Theoretical Division, T-8, Los Alamos National Laboratory, Los Alamos, NM 87545}  
\begin{abstract}
  In several classes of particle physics models -- ranging from the
  classical Majoron models, to the more recent scenarios of late
  neutrino masses or Mass-Varying Neutrinos -- one or more of the
  neutrinos are postulated to couple to a new light scalar field. As a
  result of this coupling, neutrinos in the early universe instead of
  streaming freely could form a self-coupled fluid, with potentially
  observable signatures in the Cosmic Microwave Background and the
  large scale structure of the universe.  We re-examine the constraints
  on this scenario from the presently available cosmological data and
  investigate the sensitivity expected from the Planck satellite. In
  the first case, we find that the sensitivity strongly depends on
  which piece of data is used.  The SDSS Main sample data, combined
  with WMAP and other data, disfavors the scenario of three coupled
  neutrinos at about the 3.5$\sigma$ confidence level, but also favors
  a high number of freely streaming neutrinos, with the best fit at
  5.2.  If the matter power spectrum is instead taken from the SDSS
  Large Red Galaxy sample, best fit point has 2.5 freely streaming
  neutrinos, but the scenario with three coupled neutrinos becomes
  allowed at $< 2 \sigma$. In contrast, Planck alone will exclude even
  a single self-coupled neutrino at the $4.2\sigma$ confidence level,
  and will determine the total radiation at CMB epoch to $\Delta
  N_\nu^{eff} = ^{+0.5}_{-0.3}$ ($1\sigma$ errors). We investigate the
   robustness of this result with respect to the details of Planck's
  detector. This sensitivity to neutrino free-streaming implies that
  Planck will be capable of probing a large region of the Mass-Varying
  Neutrino parameter space.  Planck may also be sensitive to a scale of
  neutrino mass generation as high as 1 TeV.

\end{abstract}

\maketitle

\section{Introduction}

Cosmology and particle physics are becoming increasingly linked, as is particularly evident in the dark sector, comprising
dark matter, dark energy and dark radiation (neutrinos or other thermalized relativistic particles).  Considerable work has been
done to discover the nature of the new particle physics underlying the
cosmological dark sector. While much remains unknown about it, the
current abundance of data, combined with more precise measurements in
the future, promise to give a clearer window into the dark.  In this
paper we consider the neutrino fraction of the dark
sector.  Cosmology already tells us a number of interesting facts
about the neutrino sector when the universe was in its infancy.  Big
Bang Nucleosynthesis (BBN), for example, places significant constraint
on the amount of radiation at $T \sim 1 \mbox{ MeV}$, translating to
an effective constraint on the total number of thermalized neutrino
species, $N_\nu^{eff} < 4.5$ \cite{Cyburt:2004yc}.  The
strongest upper bound on neutrino mass is also currently derived from
the matter power spectrum on small scales, and the current limit is $\sum m_\nu < 0.17 \mbox{ eV}$ at 95\% C.L., as given in
\cite{Seljak:2006bg}.

From a particle physics perspective, there are many well-motivated
models which generate beyond-the-standard-model neutrino physics,
potentially giving rise to signals in the early universe.  Most of
these proposals could be classified into one of the following four
broad categories: suggestions that neutrinos have (i) small but
nonzero masses \cite{Pontecorvo:1967fh,Gribov:1968kq}; (ii)
interactions with electromagnetic fields (via magnetic/transition
moments) \cite{Cisneros:1970nq,Okun:1986na}; (iii) interactions with
new heavy states \cite{Wolfenstein:1977ue}; (iv) interactions with new
light states \cite{Chikashige:1980ui,Gelmini:1980re}. As is well
known, the first of these ideas has actually proven true, giving
us the first direct discovery of physics beyond the standard model.
While it is certainly possible that there are no other observable
effects of new physics, it is intriguing to think that the neutrino
sector could hold more surprises. The pragmatic approach is to set
aside any theoretical prejudice (which has proven rather unhelpful in
the past in predicting neutrino properties) and settle the issue
experimentally. Should new neutrino interactions be discovered, the
theoretical implications would be profound.

Testing neutrino properties in terrestrial experiments has been very
challenging.
It took three decades from the first indication of the ``solar
neutrino problem" \cite{Davis:1968cp} to conclusively demonstrate
that neutrinos are massive and oscillate.
As for the other proposals on the list, progress has been even slower.
In the case of non-standard neutrino interactions mediated by novel
heavy particles, bounds from accelerator-based experiments
\cite{Berezhiani:2001rs,Davidson:2003ha}, solar
\cite{Friedland:2004pp,Guzzo:2004ue,Miranda:2004nb}, atmospheric
\cite{Friedland:2004ah, Friedland:2005vy} and beam neutrino
experiments \cite{Kitazawa:2006iq,Friedland:2006pi} remain quite weak
to this day. Bounds on the electromagnetic moments changed little over
the last 15-20 years, with the best constraints coming from cooling of
astrophysical systems such as red giants before helium flash
\cite{Raffelt:1990pj}.  Finally, while certain models of neutrino
coupling to light scalar fields (those involving weak doublet
\cite{Aulakh:1982yn,Santamaria:1987uq} or triplet
\cite{Gelmini:1980re} fields) have been ruled out by the LEP data on
the $Z$ boson width, a large class of models (involving coupling to
singlet fields) remains viable (see, {\it e.g.},
\cite{Berezhiani:1992cd}).

In this paper, we show that the experimental sensitivity to
nonstandard interactions mediated by a new light particle is about to
be significantly improved.  In this case the experimental tools are
cosmological observations rather than neutrino oscillation
experiments. The new interaction can manifest itself through large
scattering rates of the neutrinos with themselves (and the new light fields), or
through extra radiation in the early universe (through thermally
populating the new light states).  As we will show, while the present
cosmological data yields limited constraints on the scenarios in
question, the data on the Cosmic Microwave Background (CMB)
anisotropies expected in the next several years from the Planck
satellite mission (scheduled to launch in 2008) will lead to qualitative improvements.

There are many models which generate strong neutrino interactions.  A
sterile neutrino added to the standard model may have its mass
generated by a scalar field in analogy with the Higgs mechanism
(though the mass-generating scalar is not, by necessity, the Higgs
boson since the sterile neutrino carries no standard model
interactions).  That scalar will mediate neutrino interactions, and
the size of the coupling is often large enough to make them measurable
in the CMB.  Cases in point are Mass-Varying Neutrinos (MaVaNs)
\cite{Fardon:2003eh,Fardon:2005wc}, a theory of neutrino dark energy,
or a model of ``late'' neutrino masses
\cite{Chacko:2003dt,Chacko:2004cz}.  These theories feature
neutrino-scalar interactions of a size which, as we will see, may be
large enough to be measurable by the Planck satellite.  We will see
that Planck alone will be able to observe or rule out scenario with a
single interacting and two standard neutrinos at the 4.2 $\sigma$
level.

In many models, additional scalars may also become thermally
populated. 
It is possible to choose the parameters such that the scalars
recouple after the BBN era \cite{Beacom:2004yd,Chacko:2003dt}. Hence an
independent CMB constraint becomes of great interest. Though the
present constraints on $N_\nu^{eff}$ are not very strong in
comparison to the BBN constraints, we will also find that Planck
will be able to observe a single extra neutrino at approximately
2.5 $\sigma$ level.

It has been shown that in the case of interactions mediated by heavy
particles cosmology does not yield effective new bounds
\cite{Mangano:2006ar}. Why is the situation different when the new
particles are light? The basic idea can be easily stated. Coupling
neutrinos to a light scalar field can populate additional degrees of
freedom in the early universe, changing the expansion rate
as well as the evolution of inhomogeneities.
Moreover, it can drastically increase the neutrino-neutrino
interaction cross section, much more than what is possible with heavy
new particles, turning the neutrinos in the early universe
into a self-coupled fluid. The neutrino density perturbations would
evolve differently from the standard case -- coupled neutrinos would
undergo acoustic oscillations rather than stream freely in the CMB
epoch ($T\sim 1-10$ eV) -- and affect the baryon-photon fluid and dark
matter through their gravity during the epoch of radiation domination.

Our cosmological analysis is done in a rather model-independent way.
For most of the parameter space, one can simply assume that in the CMB
epoch there is a certain number of free streaming relativistic
species, $N_{FS}$ (the standard neutrinos fall into this category),
and a certain number of relativistic species forming a self-coupled
fluid, $N_{coupled}$. This approach is certainly not new
\cite{Chacko:2003dt,Bell:2005dr}.  We will denote the total number of
effective neutrinos (which includes any thermalized radiation, whether
free-streaming or coupled) as $N_\nu^{eff}$.  The results then have a
broader applicability.  For example, if one sets $N_{coupled}=0$, one
obtains a bound on the total number of the standard neutrinos.  The
constraints on the latter from the current data, as well as forecasts
for Planck, have been investigated in numerous recent studies
\cite{Bell:2005dr,Seljak:2006bg,Cirelli:2006kt,Hannestad:2006mi},
\cite{Bowenetal,Bashinsky:2003tk,Perotto:2006rj}, with somewhat
differing results. We will weigh in on this controversy.

The cosmology of neutrinos coupled to a light scalar is a very old
topic. Over the years, different aspects of it were discussed in
\cite{Georgi:1981pg,Kolb:1985tq,Raffelt:1987ah,Atrio-Barandela:1996ur},
\cite{Chacko:2003dt,Beacom:2004yd}.
Four modern data analyses are those by Hannestad
\cite{Hannestad:2004qu}, by Trotta and Melchiorri
\cite{Trotta:2004ty}, by Bell, Pierpaoli, and Sigurdson
\cite{Bell:2005dr}, and by Cirelli and Strumia
\cite{Cirelli:2006kt}. These papers only analyze the existing data
and do not forecast the reach of Planck. Moreover, once again, the
results these papers find with the available data differ.
We will reexamine this situation using
up-to-date cosmological data.

In the next Section we examine in more detail the types of neutrino
models which generate strong neutrino interactions or extra
relativistic degrees of freedom.  We then turn to the CMB
phenomenology of constraining these models.  We carry out an analysis
using all the current cosmological data, and determine the precision
with which Planck will be able to constrain neutrino interactions.  We
then apply these constraints to Majorana neutrino interactions and
consider the implications for theories of neutrino dark energy
(MaVaNs).  The details of our analysis are contained in Sects. III-V;
the Reader with interests in cosmology may wish to focus on these
Sections. Conversely, the Reader who is interested only in the
implications for neutrino models can proceed directly from Sect. II to
Sect. VI.

\section{Models of Neutrino-Scalar Interactions}
\label{sect:neutrinomodels}

As mentioned in the Introduction, the possibility of coupling the
neutrinos to a novel light scalar field has been entertained in the
literature since the early 1980's. In this Section, we will briefly
review some of the cosmologically relevant features of the many models
that were constructed. For more details, the Reader should consult,
{\it e.g.},
\cite{Chikashige:1980qk,Chikashige:1980ui,Gelmini:1980re,Georgi:1981pg,
Aulakh:1982yn,Santamaria:1987uq,Berezhiani:1992cd} and many other papers dedicated
to the subject.

 From the point of view of cosmological constraints based on the
CMB, we are interested in the physics at 1 eV energies. At
these low energies the relevant properties of the full models are
captured by effective low-energy Lagrangian terms. We may thus
try to learn about some of the common features of the models by
building this interaction ``bottom-up''.

The starting step may be to seek an interaction of the Yukawa type,
$g\phi \nu \nu$. To have the correct gauge structure, the neutrino
field must be promoted to the lepton doublet. At the level of
dimension 4 operators one can then write $L^c \boldsymbol\tau
\boldsymbol\cdot\boldsymbol\phi L$. This means the new scalar field
$\boldsymbol\phi$ is a triplet of $SU(2)$ and couples to the $Z$
boson.  If one further writes a symmetry breaking potential for
$\boldsymbol\phi$ (to obtain a massless Goldstone boson -- a Majoron),
one obtains the classical model of Gelmini and Roncadelli
\cite{Gelmini:1980re}. This model has been ruled out by the LEP
measurements of the $Z$ boson width: the triplet Majoron would
contribute an equivalent of two extra neutrino species to the width.
This argument extends to other models in which the new light scalar is
not a singlet of $SU(2)$; for instance, the Higgs doublets
\cite{Aulakh:1982yn,Santamaria:1987uq} would contribute a half of an
extra neutrino species to the $Z$ width.  As a consequence, we must
assume that the light scalar field is a Standard Model singlet.

The simplest renormalizable model that generates an effective
low-energy neutrino-scalar coupling
involves adding a right-handed sterile neutrino
$N$ to the theory and coupling a scalar $\phi$ to it.  To this end,
consider the standard see-saw mechanism. The mixing of $N$ with the
Standard Model neutrinos only enters through a Dirac neutrino mass
term, so that the neutrino Lagrangian contains
\begin{eqnarray}
   \label{eq:plainseesaw}
  {\delta\cal L} = y L H N + m_N N N.
\end{eqnarray}
The first term is of the same form as the lepton Dirac masses in the
Standard Model, while the second one is a Majorana mass term for the
right-handed (sterile) neutrino.  When the sterile neutrino is
integrated out, assuming $m_N \gg m_D\equiv y v$, only a light
Majorana neutrino remains, with mass $y^2 v^2 /m_N$.
 
We may now promote the sterile neutrino mass, $m_N$, to a dynamical
(complex) field $\phi$, in analogy to the Standard Model Higgs
mechanism, where the Dirac mass is generated by the dynamical Higgs
\footnote{Clearly, being an $SU(2)$ singlet, $\phi$ cannot be the
  Higgs}: 
\begin{eqnarray}
{\delta\cal L} = y L H N + \lambda \phi N N.
 \label{dynamicalseesaw}
\end{eqnarray}
When $\phi$ develops a VEV, $\left<\phi\right>={f}$, the sterile
neutrino gets a mass $m_N=\lambda{f}$ and can be integrated out of the
theory.  The light Goldstone mode $G$ ($\phi\equiv ({f}+\rho)
e^{-iG/{f}}$) is coupled to the light neutrinos via an effective
interaction of the form
\begin{equation}\label{eq:dim6operator2}
    \frac{y^2}{\lambda{f}} e^{i G/{f}} (L\left<H\right>)(L\left<H\right>).
\end{equation}
Clearly, at the energies of $O(1)$ eV the Higgs field in
Eq.~(\ref{eq:dim6operator2}) has no excitations, only the vacuum
expectation values (VEVs), $\left<H\right>= v $.

An alternative form of the interaction can be obtained by absorbing
$e^{i G/f}$ into the phase of $L$. The interaction then appears from
the kinetic term for $L$,
 \begin{equation}\label{eq:derivativecoupling}
    \frac{i}{f}\partial_\mu G
    \bar{L}\bar\sigma^\mu L+h.c.
\end{equation} 
This form makes manifest the derivative nature of the coupling of $G$.

We note that while the construction reviewed here is the simplest, it
is certainly not unique. In general, the scalar field need not couple
to only the sterile neutrino and could be coupled
via \emph{both} the Dirac and Majorana terms in
Eq.~(\ref{eq:plainseesaw}), {\it e.g.}, by promoting
$y\rightarrow\phi/\Lambda$, $m_N\rightarrow \Phi^2/\Lambda$
\cite{Chacko:2004cz}.

Expanding Eq.~(\ref{eq:dim6operator2}) to the first order in $G$, we
find a Majorana mass for $L$, $m_\nu=y^2 v^2/(\lambda{f})$ (where $v = \langle H \rangle$)
and an effective Yukawa (pseudoscalar) interaction of $G$ with $L$,
with the coupling strength 
\begin{eqnarray}
  \label{eq:g}
  g=\frac{y^2 v^2}{\lambda{f}^2}
  =\frac{\lambda m_D^2}{m_N^2}
  =\frac{m_\nu}{{f}}.
\end{eqnarray}
The second equality has a very simple physical interpretation:
$m_D^2/m_N^2$ gives the amount of admixture of the sterile neutrino
into the light mass eigenstate and $\lambda$ is the coupling of $g$ to
the sterile component.

In order to generate a sizable coupling $g$ between the neutrino and
scalar, there must not be too large a hierarchy between $m_\nu$ and
the scale of symmetry breaking $f$.  If the sterile neutrino
mass is at GUT scale, the coupling of $G$ is of the size $g \sim
10^{-25}$, assuming $m_\nu \sim 0.1 \mbox{ eV}$ and $\lambda \sim 1$,
which is too small to give rise to significant effects
during the CMB epoch.  On the other hand, if the scale of new physics,
$\Lambda$, is TeV scale, couplings of the size $ g \sim 10^{-13}$ may
be generated.  Though this is a very small coupling, it is large
enough to result in observable effects through decay-inverse decay of
the neutrino to scalars, which tightly couples the scalars and
neutrinos, removing neutrino free-streaming.  We return in detail to
the constraints on the coupling $g$ in
Sect.~\ref{sect:discussconstraints}.

There are many theories where a coupling $g > 10^{-13}$ is generated.
Theories of neutrino dark energy introduce interactions in the
effective theory of the form Eq.~\ref{dynamicalseesaw}, often with a
sizable coupling $g$.  In the models of dark energy originally
discussed in \cite{Frieman:1991tu} and revived in \cite{Hill:2006hj}
as neutrino dark energy, the scalar generating the dark energy is a
pNGB.  If the scale of the symmetry breaking $f$ of the scalar U(1)
symmetry is sufficiently low (TeV scale or lower), a cosmologically
interesting coupling between the scalar and neutrino may be generated.
Since the scalar is a pNGB in these models, it is protected from large
radiative corrections, and hence remains light--an advantage in any
theory of dark energy.  In a similar model of \cite{Barbieri:2005gj},
$f \sim M_{pl}$, but $m_N$ and $m_D$ are small, through extremely
small Yukawa couplings $y$ and $\lambda$ to the neutrino mass
generating fields $H$ and $\phi$.  Since $\lambda$ is very small, $g$
is also very small and there are no observable consequences in the CMB
through the neutrino-scalar coupling.  Because the sterile neutrino is
light, however, and the mixing between the active and sterile neutrino
is substantial, the sterile neutrino may become thermalized, and there
may be signals in the CMB through increased $N_{FS}$.

Another theory of dark energy may also generate interesting signals in the CMB.  In contrast to these models where the scalar field is typically
associated with a broken symmetry at a comparatively high scale $f$, MaVaNs, as
introduced in \cite{Fardon:2003eh}, place the entire sector around a
meV--this scale is chosen according to the
measured neutrino mass splittings and the dark energy scale $\delta
m_\nu^2 \sim 10^{-1} \mbox{ eV}$, $\Lambda_{DE} \sim 10^{-2.5} \mbox{ eV}$.  As a consequence,
all mass scales in the model, including the Dirac mass and the
scalar mass, lie in the sub-eV range, and the coupling $\lambda$ is
typically not too small; the small hierarchy between the
Dirac and sterile neutrinos implies a sizable coupling $g$ in many
cases.

There are many other instances where the hierarchy between the Dirac
neutrino and sterile neutrino is much smaller due to the introduction
of lighter sterile neutrinos.  Much lighter sterile neutrinos have
been considered in a wide variety of contexts, most notably perhaps in
connection with the LSND measurement, where the presence of a sterile
neutrino with mass around $1 \mbox{ eV}$ has been invoked to explain
the appearance of muon neutrinos \cite{Gonzalez-Garcia:2002dz}.  Other
models feature a keV mass sterile neutrino as the dark matter
\cite{Dodelson:1993je,Abazajian:2001nj,Shi:1998km}, sometimes with
accompanying keV mass scalars \cite{Kusenko:2006rh}, and weak scale
neutrinos associated with SO(10) GUTS, where the addition of the
neutrino is necessary for anomaly cancellation \cite{Raby:2002er}.
Various low energy see-saws have also been considered, as in
\cite{Chacko:2003dt}, where the scale $f$ is in the 50 MeV to 500 GeV
range.

We have seen that there is a broad class of models which generate
exotic neutrino-scalar interactions.  The addition of these
interactions, depending on the choices of $\lambda$, $m_D$ and $m_N$,
may be good candidates for observation in the cosmic microwave
background.  First, additional scalars may become thermalized and
increase the effective number of neutrino species.  Second, these
scalars mediate additional neutrino interactions, through scalar
mediated neutrino scattering, which could remove neutrino
free-streaming at CMB temperatures.  We now turn our attention to
studying the impact of non-standard behavior in the dark radiation
sector on CMB anisotropies.

\section{Tightly coupled neutrinos: modified evolution equations.}
\label{sect:modifiedeqns}

We summarize the relevant physical effects of dark radiation (i.e.
neutrinos) on the CMB.  The energy density in relativistic neutrinos
\footnote{Given the modern direct bounds on the neutrino mass, at
  temperatures above a few eV the neutrinos are certainly
  relativistic.}
is a fraction $\simeq N_\nu^{eff} /(N_\nu^{eff}+ 4.4 )$ of the total
radiation (freely-streaming and coupled neutrinos plus photons). Thus
during the radiation epoch the neutrino gravity is important. One
needs to consider both the effects of the neutrino background density
and that of the neutrino density perturbations.  $N_\nu^{eff}$
dictates the expansion rate of the universe and, together with matter
density $\Omega_m h^2$, controls the redshift of matter-radiation
equality, $z_{eq}$. The latter effect is more subtle. Assuming
adiabatic initial conditions, the neutrino and photon inhomogeneities
are of comparable size to begin, so the presence of neutrinos modifies
the evolution of the photon perturbations. When a perturbation of a
given size enters the horizon, the gravity of the neutrino
perturbation is comparable to the gravity of the photon perturbation.
The subsequent evolution of the two, however, are different. The
photon-baryon plasma oscillates like a compressible fluid; the
standard neutrinos, on the other hand, stream freely, quickly erasing
their density fluctuations.  Gravitational coupling between the two
means that the evolution of fluctuations in the photons could be
affected by neutrino free-streaming.  In fact, Stewart
\cite{Stewart:1972} noted back in 1972 that if this effect were large,
it would jeopardize structure formation.

Shortly thereafter, Peebles \cite{Peebles:1973} numerically solved the
problem of the coupled evolution of the neutrino and photon
perturbations. He showed that the neutrino inhomogeneities in the
standard case do indeed decay shortly after entering the horizon; in
the process, they damp photon inhomogeneities, although at a
significantly smaller level ($\sim$12\%) than anticipated by Stewart.
Much more recently, the problem was addressed by more accurate
numerical computations and analytically 
\cite{Hu:1995en,Bashinsky:2003tk} and for the CMB power spectrum the
amount of suppression (relative to the case where all the radiation is
strongly coupled) was found to be
\begin{equation}
   \label{eq:amplitudesuppression}
   \delta C_\ell/C_\ell \approx - 0.53\,\rho_{FS}/\rho_{rad},
\end{equation}
where $\rho_{FS}$ is the energy density in freely streaming radiation
and $\rho_{rad}$ is the total radiation density (free-streaming plus
strongly coupled, including photons). 

Coupling the oscillating photons to the damped neutrinos through the
gravitational potential also changes the phase of the photon
oscillations, due to a change in the speed at which information propagates in the
fluid from the presence of freely streaming neutrinos, compared to the
hypothetical case of no free-streaming neutrinos.  This effect was
clearly established in \cite{Bashinsky:2003tk}. The resulting shift of
the CMB peaks is
\begin{equation}
   \label{eq:phaseshift}
   \delta \ell \approx - 57\,\rho_{FS}/\rho_{rad}.
\end{equation}

Remarkably, both the amplitude suppression and phase shift are clearly
present in the Peebles' solution \footnote{The relevant information is
  contained in
  Table I of \cite{Peebles:1973}, which gives the
  amplitude and location of the seventh oscillation maxima for photons.}.
These effects are at the core of the physics behind the sensitivity of
CMB to neutrino free-streaming.  The amount of damping and the
phase shift change if either additional freely streaming relativistic
(``neutrino-like'') species are added or the neutrinos become
self-coupled.  In the latter case, both effects are removed: the
neutrino fluid oscillates similarly to the photon fluid (without
baryon loading).

Notice that the effect is not uniform for all CMB multipoles.
The above argument was made for modes entering the horizon
in the radiation era. For modes entering the horizon in the matter
era, the neutrino perturbations have no effect. Thus, the damping
is operational only on small scales, $l\gtrsim 200$.

In our analysis, we consider interacting neutrinos in
the tightly coupled limit. By this we mean that the
neutrinos can be approximated by a fluid for the entire range of
relevant scales, including the scales corresponding to the CMB
multipoles of $l\sim 2000$ and those measured by the large scale
structure (LSS) surveys SDSS and 2dF. The neutrino analogue of the
Silk damping scale is thus assumed to be well below $\sim O(10)$
comoving Mpc. The implications of this assumption are further
discussed in Sect.~\ref{sect:discussconstraints}.

With the above assumptions, the Boltzmann equations for the coupled
neutrinos are very simple, as discussed in \cite{Hu:1995fq}: the
standard multipole expansion for the neutrino perturbations (see
\cite{MaBertschinger}) is truncated at the level of density and
velocity perturbations. The quadrupole (shear) and higher order
moments of the perturbations are set to zero. The analogues of Eqs.
(49) or (50) in \cite{MaBertschinger} are:
\begin{itemize}
\item \emph{Synchronous gauge}
\end{itemize}
\begin{eqnarray}
   \label{eq:boltzmannsynchronous}
   \dot\delta_\nu &=& -\frac{4}{3}\theta_\nu - \frac{2}{3} \dot{h},
\nonumber\\
   \dot\theta_\nu &=&  \frac{1}{4} k^2\delta_\nu,\nonumber\\
   \sigma_\nu &=& 0.
\end{eqnarray}
\begin{itemize}
\item \emph{Conformal Newtonian gauge}
\end{itemize}
\begin{eqnarray}
   \label{eq:boltzmannNewt}
   \dot\delta_\nu &=& -\frac{4}{3}\theta_\nu + 4 \dot{\phi},\nonumber\\
   \dot\theta_\nu &=&  \frac{1}{4} k^2\delta_\nu + k^2 \psi,\nonumber\\
   \sigma_\nu &=& 0.
\end{eqnarray}
Here all the conventions are those of \cite{MaBertschinger}. The
quantity $\delta_\nu\equiv\delta\rho_\nu/\rho_\nu$ is the neutrino
density perturbation; $\theta_\nu\equiv ik^j \delta T
^{0}_{\phantom{0}j}/(\bar\rho+\bar{P})$, with the ``$\nu$'' index
assumed on the right hand side, is the neutrino velocity
perturbation; $\sigma$ is the shear (see Eq. (22) of
\cite{MaBertschinger}). The quantity $h$  is one of the two scalar
perturbations in the synchronous gauge (the one corresponding to the
trace of the scalar metric perturbation). $\phi$ and $\psi$ are the
scalar metric perturbations in the Conformal Newtonian gauge. They
coincide, up to a sign, with the gauge invariant Bardeen variables
\cite{Bardeen1980}.

\begin{figure}[htbp]
   \centering
   \includegraphics[width=0.49\textwidth]{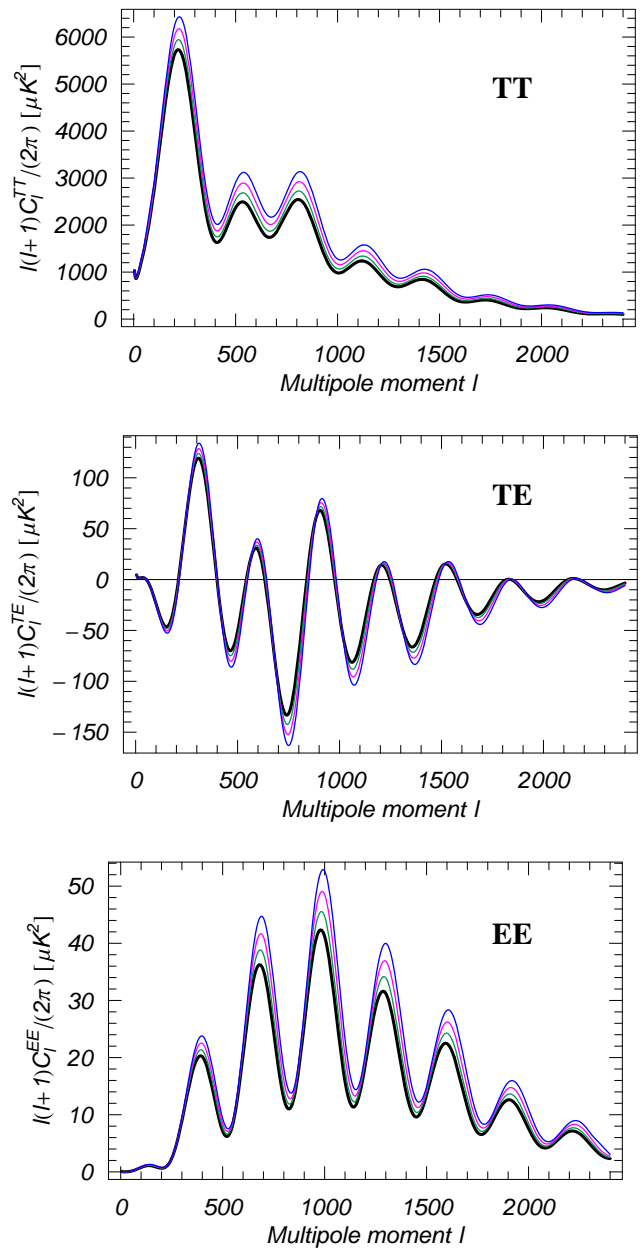}
   \caption{Effect of neutrino free-streaming on the CMB multipole spectrum.  The thickest curve is the spectrum with the best fit parameters from WMAP3; the other curves (from bottom to top) correspond to 1, 2, and 3 strongly coupled neutrinos (keeping the total number of neutrinos fixed at 3).}
   \label{fig:illustrate_coupling}
\end{figure}

The same method of truncating the multipole expansion is utilized in
the earlier analyses. Our equations agree with those of
\cite{Hannestad:2004qu,Bell:2005dr} (which are stated in the
synchronous gauge) and those of \cite{Cirelli:2006kt} (which are
stated in the Conformal Newtonian gauge). We note that the CAMB
\cite{Lewis:1999bs} code we use (also CMBFast \cite{Seljak:1996is})
employs the synchronous gauge. A slightly more general
parameterization in terms of the ``viscosity parameter'', $c_{vis}^2$
\cite{Hu:1998kj} is followed in \cite{Trotta:2004ty}.

The effect of making the neutrinos coupled, for fixed cosmological
parameters, is illustrated in Fig.~\ref{fig:illustrate_coupling}.  The
thick curve refers to the standard case of three freely streaming
neutrinos, while the other curves illustrate the effect of coupling 1,
2, and 3 neutrinos (in order of deviation from the thick curve). The
changes of the temperature (TT), polarization (EE), and the cross
correlation between them (TE) are shown. The Figure clearly exhibits
both the amplitude suppression and the phase shift effects at
$l\gtrsim 200$.

The difference in the anisotropies on scales $l\gtrsim200$ is quite
large (on average about 25-30\% for three coupled neutrinos), well
outside the errors of the current WMAP data. However, this does not
mean than the coupled neutrino scenario is already
excluded. Indeed, it must be kept in mind that there are many
cosmological (``nuisance") parameters that can be adjusted, such as the
baryon, dark matter and dark energy densities, the spectrum of
primordial fluctuations, and others. By adjusting these parameters,
it may be possible to undo most of the effect of the neutrino
self-coupling.

\begin{figure}[htbp]
   \centering
   \includegraphics[width=0.49\textwidth]{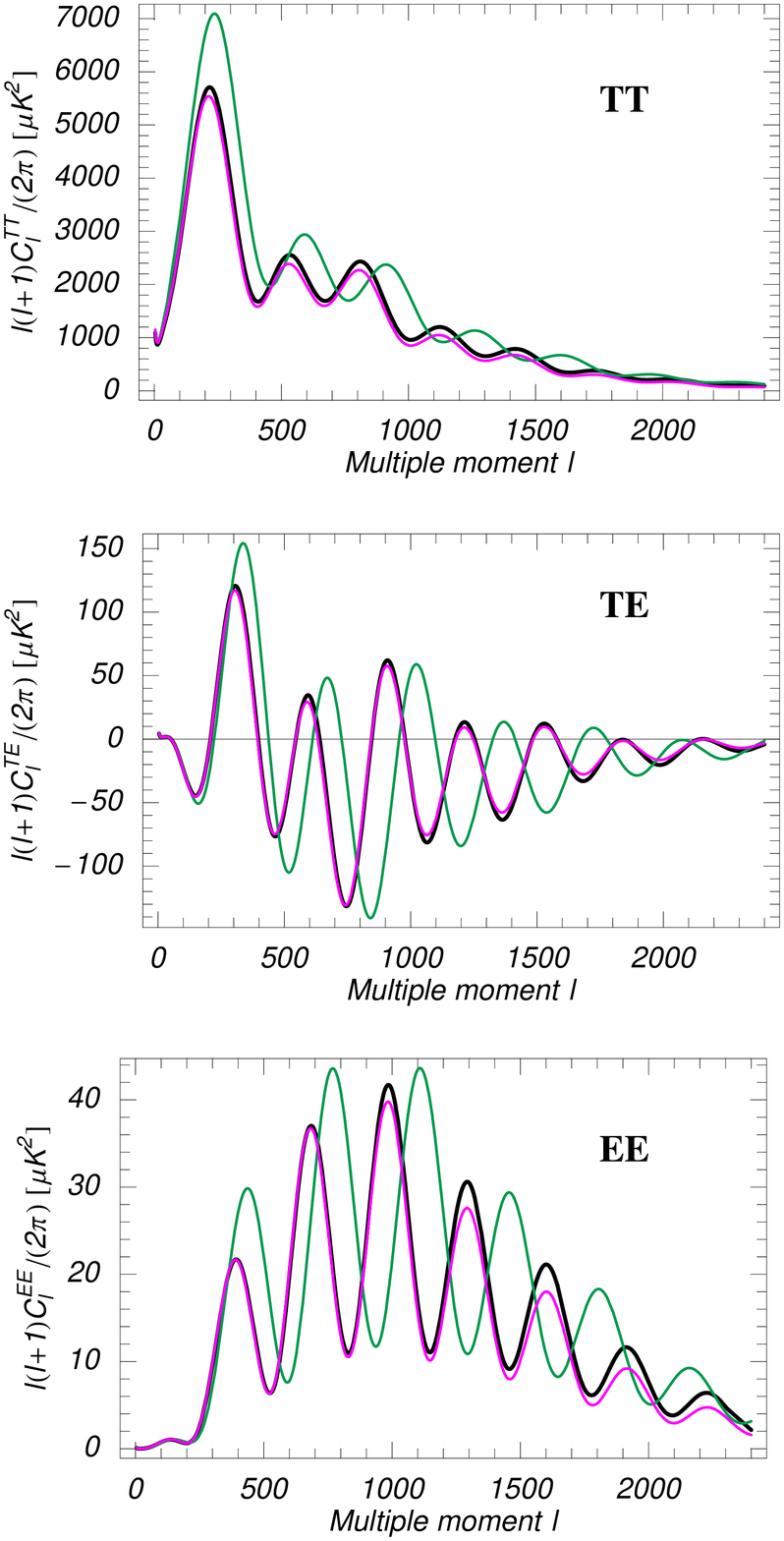}
   \caption{Effect of extra neutrinos on the CMB multipole spectrum. The central black curve is the spectrum with the best fit parameters from WMAP3, the top green curve the spectrum with 7 freely streaming neutrinos, and the lowermost magenta curve results when the total number of freely streaming neutrinos is 7, but $z_{eq}$ is kept fixed by varying $h^2$.}
   \label{fig:illustrate_Ntot}
\end{figure}

This issue of ``degeneracies'' between different parameters is of
course well known in cosmology. A simple illustration of it is given
in Fig.~\ref{fig:illustrate_Ntot} where we show changing the CMB power
spectra as one changes the total number of freely streaming neutrinos.
While simply changing $N_{FS}$ to 7 produces a large shift in the
position of the peaks (because of the faster expansion in the
radiation era as discussed above), the effect can be compensated by
changing other parameters such that the redshift of equality is
preserved \cite{Bowenetal}.  $1+z_{eq} = 4.05 \times 10^4 \Omega_m
h^2/(1+0.6905 N_\nu^{eff}/3.04) $ is fixed while varying $N_\nu^{eff}$
by scaling $h^2$ to compensate for the increase in $N_\nu^{eff}$,
while also fixing the physical baryon density $\Omega_b h^2$ and
$\Omega_m$.  Indeed, the physical quantities that are measured in CMB
are dimensionless quantities (angles on the sky), hence they depend on
the ratios of the physical densities, etc. See, {\it e.g.},
\cite{Bashinsky:2004vc} for further discussion.

Not all effects follow this simple argument. For example, the Silk
damping does not, as it involves a physical dimensionful constant, the Thompson
cross section. The faster expansion of the universe with more
neutrinos implies more Silk damping in the high multipoles of the CMB
\footnote{As the expansion rate is increased, the age of the universe
  at recombination, $t$, decreases. The sound horizon scales as $t$
  while the photon diffusion distance varies at $\sqrt{t}$. Hence, the
  damping scale becomes bigger relative to the sound horizon scale,
  which, in turn, sets the position of the peaks.}. This can be
partially compensated by adjusting the Helium fraction, as discussed
in \cite{Bashinsky:2003tk}. It should be kept in mind that this
mechanism is limited by a variety of astrophysical considerations. We
will return to this topic in Sect.~\ref{sect:physicsofsensitivity}.

The real challenge is to establish the size of the \emph{residual
  differences} of the CMB predictions in the two scenarios, after
appropriately adjusting the ``nuisance'' parameters, in comparison
with the resolution of the experiments. These residual differences
turn out to be much smaller than the differences seen in
Fig.~\ref{fig:illustrate_coupling}. This fact renders difficult
writing down a simple estimate for the predicted sensitivity of the
present and future experiments using order-of-magnitude arguments and
necessitates a detailed scan of the multidimensional parameter space.

In Sect.~\ref{sect:physicsofsensitivity} we show how well the effects
of amplitude suppression (\ref{eq:amplitudesuppression}) and phase
shift (\ref{eq:phaseshift}) can be compensated by adjusting the
cosmological parameters and how big the residual differences are. 
We will also  see which CMB multipoles are essential for
testing the neutrino sector and how robust our predictions for
Planck are. A complete analysis of this type has not been done before.

We now present the results of our numerical studies.

\section{Numerical results}

\subsection{Sensitivity of the current data}

\subsubsection{Literature overview}
\label{sect:literaturesurvey}

We begin by summarizing the bounds on the numbers of freely streaming
and self-coupled relativistic species derived in the literature.  In
the case of extra relativistic species, a number of studies have been
undertaken in the last several years. Crotty, Lesgourgues, and Pastor
found the effective number of neutrinos to lie in the interval $[1.4,
6.8]$ at $95\%$ confidence level (C.L.), by combining WMAP year one
data \cite{Bennett:2003bz} with the LSS data from 2dF
\cite{Hawkins:2002sg} and a prior on the Hubble constant. Similar
results were found by Pierpaoli \cite{Pierpaoli:2003kw} and Hannestad
\cite{Hannestad:2003xv}.  Bell, Pierpaoli, and Sigurdson
\cite{Bell:2005dr} have shown that WMAP year one data
\cite{Bennett:2003bz}, Lyman-$\alpha$ forest \cite{Viel:2004np}, and
Sloan Digital Sky Survey (SDSS) data \cite{Tegmark:2003uf} taken
together constrain the effective number of neutrinos to be
$4.9_{-2.3}^{+1.9}$ at $95\%$ C.L. Seljak, Slosar and McDonald
\cite{Seljak:2006bg} have also computed this number with more recent
data -- including the WMAP three year data (WMAP3)
\cite{Hinshaw:2006ia,Page:2006hz}, Lyman-$\alpha$ forest data
\cite{McDonald:2004xn}, SDSS measurements of the baryon acoustic
oscillations \cite{Eisenstein:2005su} (see \cite{Seljak:2006bg} for a
complete list). They find a similar result, $N_{FS} =
5.3^{+2.1}_{-1.7}$ at 95\% C.L. The WMAP team finds year-3 data
combined with large-scale structure and supernovae prefer $3.3 \pm
1.7$ neutrinos \cite{Spergel:2006hy}.  Cirelli and Strumia
\cite{Cirelli:2006kt} find $N_{FS} = 5 \pm 2$ ($2 \sigma$ errors). In
contrast, Hannestad and Raffelt \cite{Hannestad:2006mi}, constraining
neutrino mass and relativistic energy density together, determine $2.7
< N_{FS} < 4.6$ at 95\% C.L. using WMAP3 and large-scale structure
data from 2dF and SDSS.

In the case of self-interacting neutrinos, all fits are consistent
with no interactions, but with (very) different confidence levels.
Hannestad \cite{Hannestad:2004qu} concludes that in the model with
three interacting neutrinos ``it is impossible to simultaneously fit
CMB and LSS data''. From his Fig.~8 we read off a 3$\sigma$ exclusion
using only WMAP1 and 6$\sigma$ exclusion when this
data is combined with the SDSS LSS data and the HST Hubble constant
measurement (in the limit of zero neutrino mass). In contrast, Trotta and Melchiorri \cite{Trotta:2004ty}
find that the interacting neutrino scenario is disfavored at only 2.4$\sigma$.  Similarly, Bell, Pierpaoli, and Sigurdson find
$N_{coupled} < 3.0$ at 95\% C.L. Cirelli and Strumia find
the number of additional interacting neutrinos constrained to be less
than $1.3$ at $1 \sigma$ with more recent data (including WMAP3).

In short, there is quite a bit of variation between the published
results. Some of this variation could be attributed to different data
used in the calculations (for example, WMAP 1-year vs. WMAP 3-year
data releases, whether Lyman-$\alpha$ was included in the analysis or not, etc),
but some clearly must be due to the differences in the analyses
themselves. We therefore consider it well-motivated to repeat the
calculations, using the most recent data available to us. In
Sect.~\ref{sect:WMAPonly} we consider the sensitivity of WMAP3 alone,
while in Sects.  \ref{sect:WMAPplusEBCS} and \ref{sect:addLya} we add
other cosmological data.

As far as forecasting for Planck, published results on $N_{FS}$
likewise differ.  In addition, no analysis on $N_{coupled}$ has been
performed.  
Ref.~\cite{Bowenetal} finds that Planck will be able to
constrain the total number of freely streaming neutrino species  
to $\Delta N_{FS}=0.24$ (1$\sigma$ error)
using temperature information only and a sky coverage $f_{\rm sky}=0.5$.
In contrast, Ref.~\cite{Bashinsky:2003tk} finds
using Planck temperature information only $\Delta N_{FS}=0.6$
even with a more optimistic sky coverage $f_{\rm sky}=0.8$.
Both studies employ the same
technique (the Fisher matrix analysis). Finally,
Ref.~\cite{Perotto:2006rj} investigated the bounds on $N_{FS}$ using
both the Fisher analysis and Markov Chain Monte Carlo (MCMC). The
results are quite intriguing: while the Fisher analysis yields $\Delta
N_{FS}=0.26(0.27)$ with (without) lensing, the MCMC method yields very
different results depending upon whether lensing is assumed -- with
lensing the result is consistent with the Fisher analysis, while
without lensing $\Delta N_{FS}=0.46$. The forecast for Planck's
sensitivity to the number of self-coupled neutrinos to the best of our
knowledge has not been done. We perform a combined analysis of
Planck's sensitivity to freely streaming and self-coupled neutrino
species in Sect.~\ref{sect:Planckonly}. We also investigate, in
Sect.~\ref{sect:PlanckplusEBCS}, a potential impact of combining
Planck with other cosmological data.

\subsubsection{WMAP3 alone}
\label{sect:WMAPonly}

\begin{table*}[t]
\begin{tabular}{|c||c|c||c|c|c|c|c|c||c|c|c|c|c|} \hline
$(N_{FS}, N_{coupled})$& $\delta\chi^2$ & C.L.& $\Omega_b h^2$ &  $\Omega_c h^2$ & $\theta$ &
$\tau$ & $n_s$ & $\log[10^{10} A_s]$ & $\Omega_\Lambda$ & Age/GYr &
$\Omega_m$ & $z_{re}$ & $H_0$
\\ \hline
$(3,0)$& -- & -- &  $0.02216$ & $0.10519$ & $1.0394$ & $0.089$ &
$0.953$ & $3.02$ & $0.76$ & 13.75 & 0.24 & 11.3 & 72.8
\\ \hline
$(2,1)$& 0.2 & $0.1\sigma$ &  $0.02249$ & $0.10478$
& $1.0423$ & $0.078$ & $0.927$ & $2.93$ & $0.77$ & 13.66 & 0.23 &
10.2 & 73.8
\\ \hline
$(1, 2) $  &0.4 & $0.2\sigma$ &
$0.02323$ & $0.10367$ & $1.0472$ & $0.080$ & $0.907$ & $2.86$ &
$0.78$ & 13.43 & 0.22 & 10.1 & 76.5
\\ \hline
$(0,3) $& 1.4 & $0.7\sigma$ & $0.02436$ & $0.09828$ & $1.0542$ &
$0.095$ & $0.897$ & $2.80$ & $0.82$ & 13.06 & 0.18 & 11.0 & 83.0
\\\hline
$(1, 0) $& 0.6 & $0.3\sigma$  &  $0.02201$ &
$0.07183$ & $1.0521$ & $0.091$ & $0.914$ & $2.88$ & $0.77$ & 15.82 &
0.23 & 10.0 & 64.1
\\ \hline
$(5, 0) $& 0.6 & $0.3\sigma$  &  $0.02187$ & $0.13921$ & $1.0312$ & $0.088$ & $0.968$ &
$3.08$ & $0.74$ & 12.40 & 0.26 & 12.2 & 79.3\\ \hline
\end{tabular}
\caption{The best fit values of $-log(likelihood)$ (second column)
and the
  corresponding shift of the best fit parameters as a function of the
  number of coupled neutrinos. The last five columns list derived
  parameters. The initial spectrum is defined with a pivot point
  $k_{piv}=0.05$ Mpc$^{-1}$. The fit is to the WMAP3 dataset only.}
\label{table:bestfitWMAP3only}
\end{table*}

As a first calculation, we investigate how sensitive WMAP is by itself
to neutrino self-coupling. We do this by fitting WMAP3 \cite{Hinshaw:2006ia,Page:2006hz}, varying a set of cosmological
parameters (to be specified shortly) under four different assumptions:
(i) three standard free-streaming neutrinos; (ii) one neutrino
coupled, two free-streaming; (iii) two neutrinos coupled, one
free-streaming; (iv) all three neutrinos coupled. We then compare the
goodness of fit at the corresponding best-fit points. We also explore
the sensitivity of WMAP to the total number of (standard) neutrinos,
in a similar way. For this, we consider two more scenarios: one with
the total neutrino number equal to five and another with a single
neutrino flavor.

The fitting is done using the MCMC code COSMOMC \cite{Lewis:2002ah},
with the CAMB code \cite{Lewis:1999bs} modified by us to include both
freely streaming and tightly self-coupled neutrinos. The MCMC method
is by now a standard tool in cosmology, used for both data analysis
and forecasting. It is employed by the WMAP
\cite{Spergel:2003cb,Spergel:2006hy} and Planck \cite{PlanckBlueBook}
teams, as well as many other groups (in particular, among the papers
reviewed in Sect.~\ref{sect:literaturesurvey},
\cite{Trotta:2004ty,Bell:2005dr,Seljak:2006bg,Perotto:2006rj} use
MCMC). With the MCMC method, the likelihood function is mapped out in
a multidimensional region of parameters around its maximum. As a
result, one gets the location of the best-fit point, with the
corresponding likelihood characterizing the goodness of fit, as well
as the allowed ranges of the parameters. One need not make {\it a
  priori} assumptions on the functional form of the likelihood
function, although choosing the parameterization in such a way that
the posterior distributions are approximately Gaussian and there are
no strong correlations saves computer time \cite{Lewis:2002ah}
\footnote{In principle, one does need to ensure that there are no
  other minima of the likelihood, well separated from the region where
  the sampling is done.}.

In this Subsection, we choose the following six cosmological
parameters: the physical densities of baryons, $\Omega_b h^2$, and
dark matter, $\Omega_c h^2$, the ratio of the approximate sound
horizon to the angular diameter distance $\theta$, the optical depth
to the last scattering surface $\tau$, and the primordial spectrum of
the scalar curvature perturbations, characterized by the spectral
index $n_s$ and the power $A_s$ on a preset (``pivot") scale (taken in
this calculation to be $k_{piv}=0.05$ Mpc$^{-1}$). We assume that the
Universe is flat, there is no running of the spectral index, the
Helium fraction is fixed at $Y_{\rm He}=0.24$ and the
dark energy is a cosmological constant ($w=-1$). As we will see, in
the case of WMAP3, the six parameters we vary contain the necessary
degeneracies with the neutrino coupling.

The results are tabulated in Table \ref{table:bestfitWMAP3only}.  The
first row corresponds to the standard scenario, $N_{FS} = 3,
N_{coupled} = 0 $. The next three rows correspond to coupling one, two,
or three of the neutrino species, while keeping the total number of
neutrinos at three. Finally, in the last two rows, we vary the total
number of neutrino species, assuming they are all freely streaming.

The second column shows the difference between the $\chi^2$ of the
best fit in a given scenario and the corresponding quantity in the
standard scenario (first row). The third column shows the
corresponding confidence level (C.L.) \footnote{Assuming 2 degrees of
  freedom, $N^{stream}$ and $N^{coupled}$.}. The next six columns in
the Table show the values of the best-fit parameters for each
scenario. The last five columns show the values of derived parameters:
the cosmological constant $\Omega_\Lambda$, the
age of the Universe, the matter fraction $\Omega_m$, the red-shift of
ionization $z_{re}$ and finally the Hubble constant, $H_0$. To
compensate for the effect of neutrino coupling, $H_0$ increases, while
$\Omega_m$ and the spectral index decrease. To compensate for
additional neutrino species, all three quantities increase. In the
latter case, it is easy to check that the best-fit parameters shift in
such a way that the redshift of matter-radiation equality, $z_{eq}$,
is preserved, as discussed in Sect.~\ref{sect:modifiedeqns}.  $A_s$ also increases with $N_{FS}$ to compensate for the reduction of perturbations on all scales. 

From the second column, we see that the effects of the neutrino
coupling as well as the variation of the total number of neutrinos
are nearly perfectly compensated. The quality of the fits is
virtually the same in each of the six cases.
Therefore, WMAP by itself cannot distinguish between these scenarios.
This conclusion is consistent with the findings of the analysis by the
WMAP collaboration, which considers the sensitivity of the experiment
to the number of freely streaming neutrino species
\cite{Spergel:2006hy}. It differs from the findings of
\cite{Hannestad:2004qu} where a 3$\sigma$ exclusion of the
($N_{FS}=0,N_{coupled}=3$) scenario was claimed. Clearly, our
conclusion relies on the ability of the code to find the set of
parameters that give the most complete compensation of the effects of
modifying the neutrino sector. We see that the MCMC method
accomplishes this task well.

\subsubsection{WMAP3 plus large scale structure (2dF, SDSS), HST, SN Ia. }
\label{sect:WMAPplusEBCS}

\begin{table*}[t]
\begin{tabular}{|c||c|c||c|c|c|c|c|c|c|c||c|c|c|c|c|c|} \hline
$(N_{FS}, N_{coupled})$
& $\delta \chi^2$ & C.L.
& $\Omega_b h^2$ &  $\Omega_c h^2$ & $\theta$ &
$\tau$ & $Y_{\rm He}$ & $n_s$ & $n_{\rm run}$ & $\log[10^{10} A_s]$ &
$\Omega_\Lambda$ & Age/GYr & $\Omega_m$ & $z_{re}$ & $H_0$& $\sigma_8$
\\ \hline
$(3, 0) $
& -- & -- &
$0.02259$ & $0.11083$ & $1.0418$ & $0.073$ & 0.244 & $0.985$ & 0.018 & $3.03$
& $0.74$ & 13.70 & 0.26 &  9.9 & 71.4 & 0.791
\\ \hline
$(2,1) $& 4.0 & $1.5\sigma$ &
$0.02340$ & $0.11703$ & $1.0478$ & $0.079$ & 0.236 & $0.956$ & 0.062 & $3.00$
& $0.73$ & 13.47 & 0.27 &10.3 & 71.8 & 0.793
\\ \hline
$(1, 2) $
& 8.8 & $2.5\sigma$ &
$0.02390$ & $0.11993$ & $1.0506$ & $0.074$ & 0.242 & $0.926$ & 0.008 & $2.93$
& $0.72$ & 13.36 & 0.28 &  9.8 & 72.0 & 0.770
\\ \hline
$(0, 3) $
& 15.6 & $3.5\sigma$ &
$0.02406$ & $0.12599$ & $1.0563$ & $0.074$ & 0.242 & $0.888$ &-0.017 & $2.89$
& $0.71$ & 13.18 & 0.29 &  9.9 & 72.5 & 0.776
\\ \hline
$(1, 0) $
& 12.3 & $3.1\sigma$ &
$0.02296$ & $0.08411$ & $1.0586$ & $0.083$ & 0.242 & $0.958$ & 0.015 & $2.98$
& $0.71$ & 15.59 & 0.29 &  9.8 & 61.2 & 0.735
\\ \hline
$(5, 0) $
& -0.3 & -- &
$0.02210$ & $0.13999$ & $1.0315$ & $0.090$ & 0.241 & $0.986$ & 0.006 & $3.10$
& $0.74$ & 12.37 & 0.26 & 12.4 & 79.4 & 0.850\\ \hline
\end{tabular}
\caption{The same as Table~\protect{\ref{table:bestfitWMAP3only}},
but including WMAP3, SDSS Main and LRG data samples, 2dF, HST, and
SN Ia datasets. The last column
is the derived best fit value of $\sigma_8$.}
\label{table:bestfitWMAP3everythingbutLya}
\end{table*}

\begin{figure*}[htbp]
  \centering
  \includegraphics[width=0.77\textwidth]{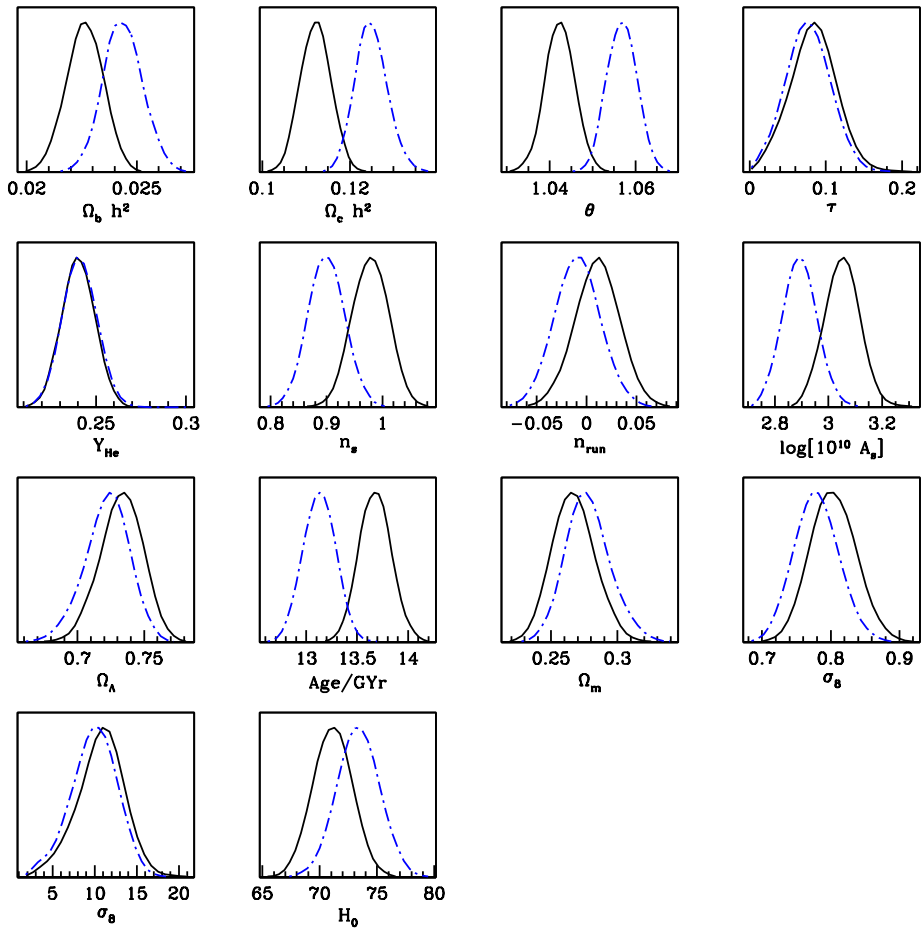}
  \caption{The shift of the best-fit parameters as a result of
    neutrino coupling. The dashed-dotted curves correspond to the scenario of
    coupled neutrinos, while the solid ones refer to the standard
    (free streaming) neutrinos. In both cases, the fits are to the
    datasets from WMAP3,
    SDSS, 2dF, HST, and SN Ia.}
  \label{fig:mpk_HST_SN_ref_vs_3coupled}
\end{figure*}

\begin{figure*}[htbp]
  \centering
  \includegraphics[width=0.77\textwidth]{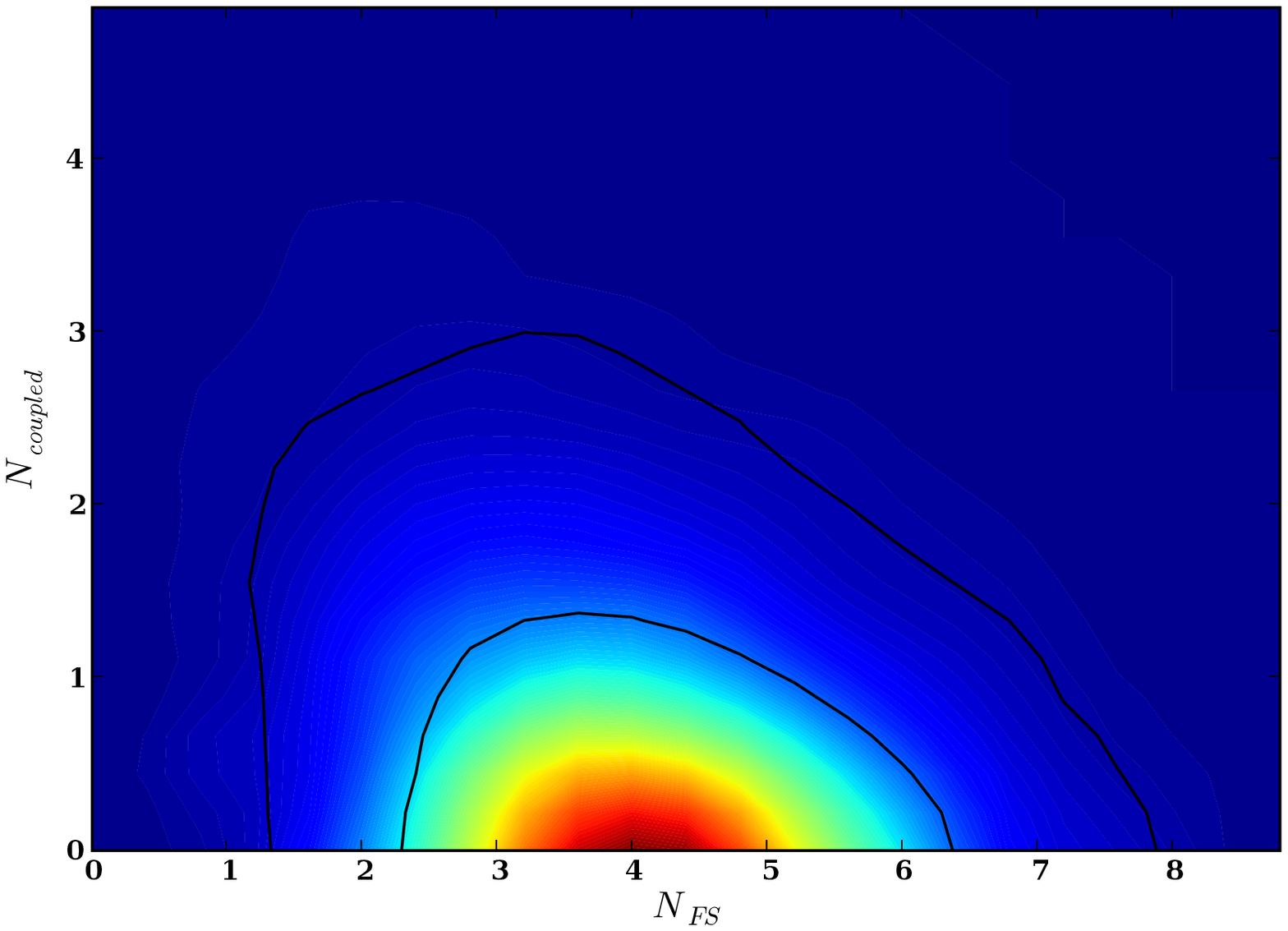}
  \caption{The combined reach of all available data (excluding Lyman-$\alpha$) on the number of 
    free-streaming neutrinos, $N_{FS}$, and the number of coupled
    neutrinos, $N_{coupled}$: WMAP3 + SDSS (LRG) + SDSS (Main) + 2dF +
    HST + SNIa.  The solid contours indicate 1 and 2$\sigma$ C.L.  }
  \label{fig:WMAP_EBCS}
\end{figure*}

We now explore if the addition of the other presently available
cosmological data can lift the degeneracy of WMAP3. In this analysis,
we include the data on the large scale structure (LSS) from the 2dF
\cite{Cole:2005sx} and the SDSS survey -- both the Main
\cite{Tegmark:2003uf} and the recently released Large Red Galaxy (LRG)
\cite{Tegmark:2006az} data samples.  We also use the Type Ia supernova
data from the SNLS collaboration \cite{Astier:2005qq}, as well as the
Hubble constant measurements from the Hubble Space Telescope (HST)
\cite{Freedman:2000cf}, in the form of a Gaussian prior $h=0.72\pm
0.8$.

We vary eight parameters, six as described in the previous
Subsection, plus two more, the helium fraction, $Y_{\rm He}$ and
the running of the spectral index, $n_s$. These two parameters are
added in anticipation of the analysis for Planck, where their
roles are important. With the present day data, as we will see,
they do not make much of a difference. For the helium fraction, we
use a gaussian prior $Y_{\rm He} = 0.24 \pm 0.009$ motivated by
observational bounds \cite{Olive:2004kq}.  We also impose a rather
generous hard prior, $0.15 < Y_{\rm He} < 0.3$. The values greater
than 0.3 would be problematic for the solar model (since the input
value into the solar model cannot be \emph{less} than the
primordial value) \cite{Ciacio:1996mk,Bahcall:2000nu}.

We begin by performing a combined fit to all the data. Results for
several specific scenarios are shown in
Table~\ref{table:bestfitWMAP3everythingbutLya}. We find that the
degeneracies of the WMAP3-only analysis are broken by the additional data.
In particular, relative to the standard case of three freely streaming
neutrinos, the case with all three neutrinos coupled is disfavored at
3.5$\sigma$, while two and
one coupled neutrinos are disfavored at 2.5$\sigma$
and 1.5$\sigma$, respectively. Assuming no self-coupled
neutrinos, the scenario with a single
freely streaming neutrino is disfavored at 3.1$\sigma$, while
the scenario with five freely streaming neutrinos gives a fit which is
just as good as the standard one.

The corresponding best-fit parameters are tabulated in
Table~\ref{table:bestfitWMAP3everythingbutLya}. One can see that the
general trends are similar to what was observed with WMAP3 only: as
more neutrinos are coupled, the fit prefers larger values of the
Hubble constant and smaller values of the spectral index $n_s$. The
shifts are graphically illustrated in
Fig.~\ref{fig:mpk_HST_SN_ref_vs_3coupled}. The standard free
streaming neutrinos are shown in solid curves, while the scenario of
three self-coupled neutrinos is shown with the dashed-dotted curves.

From the point of view of particle physics one may be interested to
know how the number of allowed self-coupled neutrino species varies
with the total number of neutrinos. As mentioned earlier, one can
easily imagine models in which additional neutrino-like degrees of
freedom are populated after the time of the BBN but before CMB (see,
{\it e.g.}, \cite{Chacko:2004cz}). In Fig.~\ref{fig:WMAP_EBCS} we
present the allowed region in the $(N_{FS}, N_{coupled})$ plane. The
plot was obtained by running MCMC in the ten-dimensional space of
parameters $(N_{FS}, N_{coupled},\Omega_b h^2, \Omega_c h^2, \theta,
\tau, Y_{\rm He}, n_s, n_{\rm run}, \log[10^{10} A_s])$, assuming a $\Lambda$CDM universe, and
marginalizing over all but the first two parameters.  From the Figure,
we see that scenarios with no freely streaming neutrino are clearly
disfavored, whether the number of self-coupled neutrinos is zero or
three. At the same time, the constraint on $N_{coupled}$ is relaxed if
in addition to the coupled neutrinos there are also $\sim3-5$ freely
streaming ones. In fact, even with standard neutrinos the fit actually
prefers the total number of freely streaming neutrinos to be greater
than three: the best fit is achieved for $N_{FS}=3.7$,
$N_{coupled}=0$.  This curious result
warrants further investigation.

\begin{figure*}[htbp]
  \centering
  \includegraphics[width=0.77\textwidth]{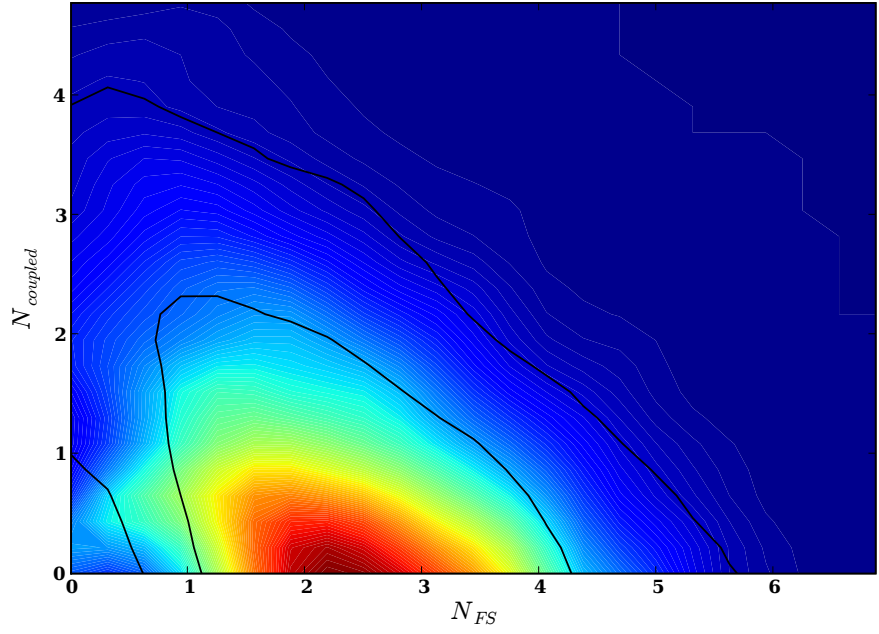}
  \caption{Sensitivity of WMAP3 + SDSS (LRG) + 2dF + HST + SNIa.  
    The removal of the SDSS Main sample lowers the best fit point 
    for $N_{FS}$ and signficantly weakens the constraints on a
    scenario with $N_{coupled} = 3$ and $N_{FS} = 0$.
   }
  \label{fig:WMAP_LRG}
\end{figure*}

It turns out that the piece of data responsible for favoring large
$N_{FS}$ is the Main data sample from SDSS. With it removed (i.e.,
with the matter power given by the LRG SDSS dataset and 2dF dataset),
the fit changes dramatically, as shown in Fig.~\ref{fig:WMAP_LRG}. The
new fit has a slight preference for values of $N_{FS}$ \emph{less}
than three: the best fit lies at $N_{FS}=2.5$, $N_{coupled}=0.1$.

\begin{figure*}[htbp]
  \centering
  \includegraphics[width=0.77\textwidth]{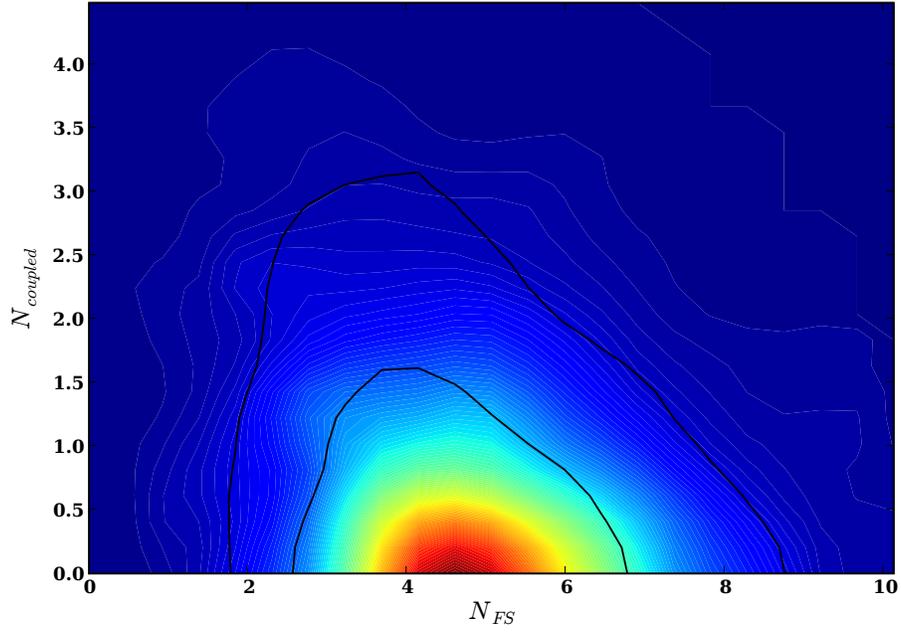}
  \caption{Sensitivity of WMAP3 + LSS from SDSS and 2dF + HST + SNIa.
    Compared to the previous figure, LRG data has been replaced with 
    SDSS Main sample.  SDSS main prefers larger $N_{FS}$ compared to 
    LSS from LRG data sample.
    }
  \label{fig:WMAP3_allbuthLya_scan}
\end{figure*}

Conversely, if we remove the LRG data from the fit, the best fit
point moves all the way to $N_{FS}=5.2$, $N_{coupled}=0.1$, as can
be seen in Fig.~\ref{fig:WMAP3_allbuthLya_scan}. Marginalizing
over all other parameters including $N_{coupled}$ yields $N_{FS}=
4_{-1.3}^{+2.2}$. Thus, it may be too early to conclude that the
SDSS LRG and Main samples are consistent with each other. At least
as implemented in COSMOMC, they pull the best-fit value of
$N_{FS}$ in different directions, with the LRG sample favoring the
standard values.

Our results show that if one chooses to rely on the LRG sample, one
finds much less sensitivity to neutrino self-coupling: the point
$N_{FS}=0$, $N_{coupled}=3$ lies inside the 2 $\sigma$ contour, while
$N_{FS}=1$, $N_{coupled}=2$ lies on the 1 $\sigma$ contour.  Thus,
at present
the bounds on coupled neutrinos from the global fit should perhaps be
taken with caution.

\subsubsection{Adding the Lyman-$\alpha$ data. }
\label{sect:addLya}

We also repeat the analysis including the Lyman-$\alpha$ dataset
in the fit. This dataset has been somewhat controversial
\footnote{Kev Abazajian, private communications.}. Our main reason
for doing this calculation is to get an idea about the additional
sensitivity that can be gained from this dataset, and also to
further check consistency with the published analyses
where this data is used.

We find that the overall sensitivity to coupled neutrinos is
somewhat increased with the addition of this piece of data. Relative
to the standard case of three free-streaming neutrinos, the case
with a single coupled neutrino is disfavored at 1.8$\sigma$,
the case with two coupled neutrinos is disfavored at 3.1$\sigma$
 and, finally, the case with three coupled neutrinos is
disfavored at 4.1$\sigma$. In other words, the addition of the
Lyman-$\alpha$ data does bring further improvements in sensitivity,
though not very large ones. The Lyman-$\alpha$ dataset does favor
large values of $N_{FS}$, just like the SDSS Main sample, as also
observed in  \cite{Bell:2005dr}, \cite{Seljak:2006bg}. The comments at
the end of the last Subsection apply here as well.

\subsection{Sensitivity of Planck}

\subsubsection{Planck only}
\label{sect:Planckonly}

As already mentioned, the situation will improve markedly with the
expected data from the Planck satellite. In this subsection, we
describe the result of our MCMC analysis of Planck's sensitivity.

We generate mock data for Planck in the \verb|all_l_exact| data format
of COSMOMC, using the best-fit point of WMAP3 as the ``true'' (seed)
value and assuming flat Universe, $w=-1$ for the cosmological
constant, and $N_{FS}=3.04$, $N_{coupled}=0$ for the numbers of
neutrinos. The characteristics of the Planck detector (the beam size
and the noise levels for temperature and polarization measurements)
are given in \cite{PlanckBlueBook}. The relevant measurements will be
done in three frequency channels of the High Frequency Instrument
(HFI), 100, 143, and 217 GHz. In the literature
\cite{Bowenetal,Bashinsky:2003tk}, the analyses have been done
assuming an effective single channel for Planck.
Ref.~\cite{Bashinsky:2003tk} in particular uses the sky coverage of
$f_{sky}=0.80$, the detector noise for temperature $w_T^{-1/2} = 40
\mu\mbox{K}'$, and for polarization $w_P^{-1/2} = 56
\mu\mbox{K}'$ and the beam size $\theta_b = 7'$ (see
Sect.~\ref{sect:Fisher} for the definitions of these quantities). We
have checked that these effective numbers (especially the value of
$w_P^{-1/2}$) are in reasonably good agreement with these
three-channel parameters of \cite{PlanckBlueBook} \footnote{Table 1.3
  of \cite{PlanckBlueBook} lists \emph{relative} sensitivities of the
  three channels, which need to be multiplied by $T_{CMB}$ to obtain
  $w_{T,P}^{-1/2}$.}. The only substantial difference is that we believe it
is more appropriate to take $f_{sky}=0.65$ \footnote{The function
  \textsf{ChiSqExact} of COSMOMC is consistent with the definitions of
  Sect.~\ref{sect:Fisher} up to the replacement $f_{sky}^2\rightarrow
  f_{sky}$. To correct for that, we feed the value
  $f_{sky}=\sqrt{0.65}$ into COSMOMC.}.

Our analysis here parallels that of the last Subsection. We first
fit the mock data in the several specific scenarios considered early,
namely, varying the numbers of self-coupled neutrinos keeping the
total neutrino number at three, and then varying the number of standard
neutrinos assuming no self-coupled neutrinos. We compare the quality
of the best fits in each case and observe how the cosmological
parameters change to compensate for the effects of neutrino coupling.
For our second analysis, we perform a global fit in the space of
parameters $(N_{FS}, N_{coupled},\Omega_b h^2, \Omega_c h^2, \theta,
\tau, Y_{\rm He}, n_s, n_{\rm run}, \log[10^{10} A_s])$. We then
marginalize over the last 8 parameters to obtain the allowed region in
the space of $N_{FS}, N^{coupled}$.

\begin{table*}[t]
\begin{tabular}{|c||c|c||c|c|c|c|c|c|c|c||c|c|c|c|c|} \hline
$(N_{FS}, N_{coupled})$
& $\delta \chi^2$ & C.L.
& $\Omega_b h^2$ &  $\Omega_c h^2$ & $\theta$ &
$\tau$ & $Y_{\rm He}$ & $n_s$ & $n_{\rm run}$ & $\log[10^{10} A_s]$ &
$\Omega_\Lambda$ & Age/GYr & $\Omega_m$ & $z_{re}$ & $H_0$
\\ \hline
$(3, 0) $
& -- & -- &
$0.02246$ & $0.10673$ & $1.0417$ & $0.089$ & 0.243 & $0.952$ & 0.001 & $3.03$
& $0.76$ & 13.68 & 0.24 & 11.3 & 72.9
\\ \hline
$(2,1) $&21.1 & $4.2\sigma$ &
$0.02304$ & $0.10485$ & $1.0460$ & $0.089$ & 0.262 & $0.948$ & 0.016 & $2.97$
& $0.78$ & 13.49 & 0.22 &11.3 & 75.5
\\ \hline
$(1, 2) $
&85.5 & $>8\sigma$ &
$0.02356$ & $0.10273$ & $1.0503$ & $0.088$ & 0.277 & $0.943$ & 0.030 & $2.89$
& $0.79$ & 13.31 & 0.21 & 11.1 & 78.3
\\ \hline
$(0, 3) $
&197.5 & $>8\sigma$ &
$0.02403$ & $0.10097$ & $1.0542$ & $0.088$ & 0.281 & $0.932$ & 0.039 & $2.81$
& $0.81$ & 13.11 & 0.19 & 11.0 & 81.5
\\ \hline
$(1, 0) $
& 91.7 & $ >8\sigma$ &
$0.02139$ & $0.08096$ & $1.0495$ & $0.088$ & 0.298 & $0.890$ &-0.017 & $2.95$
& $0.70$ & 16.06 & 0.30 & 11.2 & 58.9
\\ \hline
$(5, 0) $
& 32.9 & $5.4\sigma$ &
$0.02285$ & $0.13327$ & $1.0358$ & $0.093$ & 0.162 & $0.975$ & 0.006 & $3.08$
& $0.78$ & 12.16 & 0.22 & 11.4 & 84.2
\\ \hline
\end{tabular}

\caption{The same as in Table~\protect{\ref{table:bestfitWMAP3only}},
  but now fitting to the mock Planck dataset.}
\label{table:bestfitPlanckonly}
\end{table*}

\begin{figure*}[htbp]
  \centering
  \includegraphics[width=0.77\textwidth]{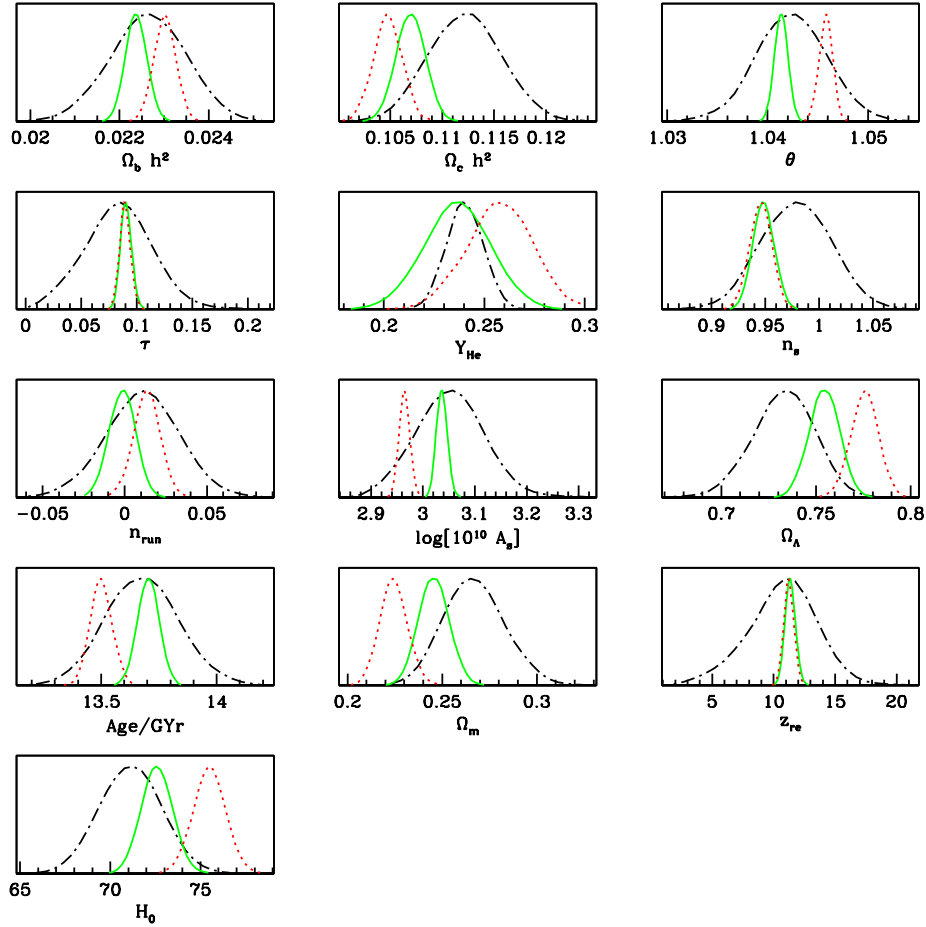}
  \caption{The improvement of sensitivity of Planck (solid) over the current
    data (dashed), assuming the standard scenario of three freely
    streaming neutrinos. The dotted curve illustrates how the
    parameters measured at Planck would shift assuming
    a single neutrino species is coupled. The current dataset is
    comprised of the data from WMAP3,
    SDSS Main and LRG, 2dF, HST, and SN Ia, plus the observational
    prior on $Y_{\rm He}$ (see text for details).}
  \label{fig:Planck_ref_vs_1coupled}
\end{figure*}

The results of the first set of calculations are tabulated in
Table~\ref{table:bestfitPlanckonly}. The trends in the parameter
shifts are similar to the case of WMAP3 -- both $n_s$ and $A_s$
decrease while the physical baryon density increases as more neutrinos
are coupled -- but with Planck the shifts are much constrained.  
Moreover, the degeneracies are very efficiently broken and the quality
of the best fits is significantly poorer in the coupled cases relative
to the standard case. The scenario of a single coupled neutrino
($N_{FS}=2$, $N_{coupled}=1$) is disfavored at 4.2$\sigma$
relative to the standard one. The scenario with ($N_{FS}=5$,
$N_{coupled}=0$) is disfavored at 5.4$\sigma$, while the other
scenarios in the Table are ruled at greater than 8$\sigma$.
Clearly, Planck's sensitivity will be dramatically better than that of
the current data.

The situation is graphically illustrated in
Fig.~\ref{fig:Planck_ref_vs_1coupled}, where the solid curves show the
expected measurements at Planck and the dashed curves are those
obtained with the combined current data (see
Sect.~\ref{sect:WMAPplusEBCS}) -- both under assumption of three
standard freely streaming neutrinos. The dotted curves show how the
best-fit parameters shift if one instead fits the data under the
assumption of a single coupled neutrino ($N_{FS}=2$, $N_{coupled}=1$).
Clearly, Planck by itself will have errors that are significantly
smaller than those of today's experiments combined.

\begin{figure*}[htbp]
  \centering
  \includegraphics[width=0.77\textwidth]{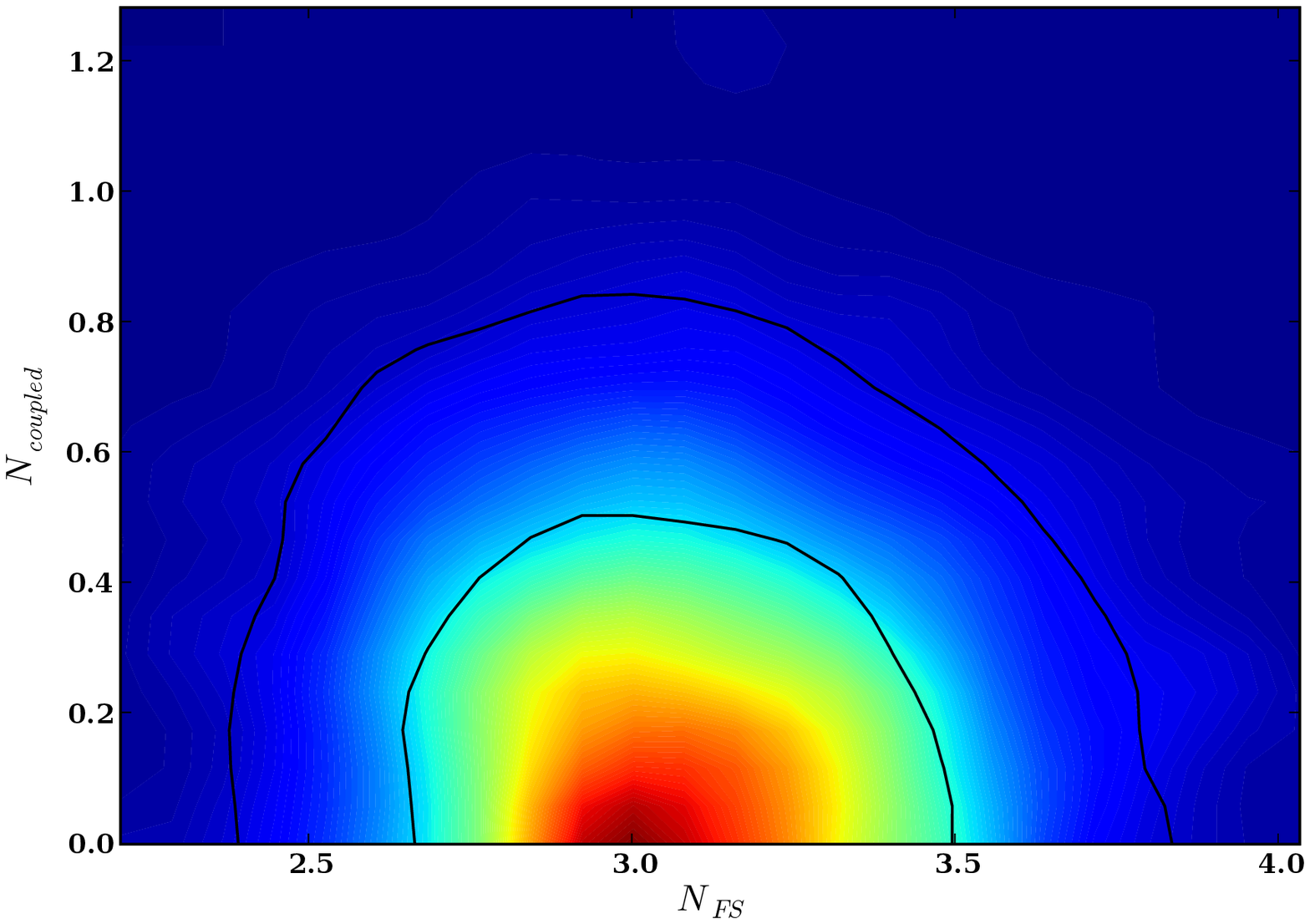}
  \caption{Expected sensitivity of Planck alone to $N_{FS}$, $N_{coupled}$. }
  \label{fig:Planck_scanNu_coupled_2d}
\end{figure*}

The results of the second calculation show that this level of
sensitivity persists for any direction in the $(N_{FS},
N_{coupled})$ plane. The sensitivity contours in this plane are
depicted in Fig.~\ref{fig:Planck_scanNu_coupled_2d}.  We wish to
stress two main results of this analysis: (i) the sensitivities of
Planck to $N_{coupled}$ and $N_{FS}$ are similar, and (ii) for
no direction in the plane $(N_{coupled}, N_{FS})$ is there is a
degeneracy between the two parameters. We will return again to the
last point in Section \ref{sect:Fisher}.

Our prediction for Planck's sensitivity to the number of freely
streaming neutrinos is asymmetric, $\Delta N_\nu = ^{+0.5}_{-0.3}$.
The lower error roughly agrees with that of \cite{Bashinsky:2003tk},
when the choice of higher $f_{sky}$ is taken into account, and
\cite{Perotto:2006rj} obtained with the Fisher matrix method. It is in
reasonable agreement with what \cite{Perotto:2006rj} finds with the
MCMC method assuming no lensing of the CMB. We have investigated the
effect of lensing on the sensitivity and we do not find the large
effect of \cite{Perotto:2006rj}. Instead we find the bounds remain
similar, slightly weaker for $N_{FS}$, slightly stronger for
$N_{coupled}$.  Notice that \cite{Bashinsky:2003tk}, considering Planck's
sensitivity to $N_{FS}$ with the Fisher analysis, also finds the effect of
lensing small.

\subsubsection{Planck plus  other cosmological data}
\label{sect:PlanckplusEBCS}

\begin{figure*}[htbp]
  \centering
  \includegraphics[width=0.77\textwidth]{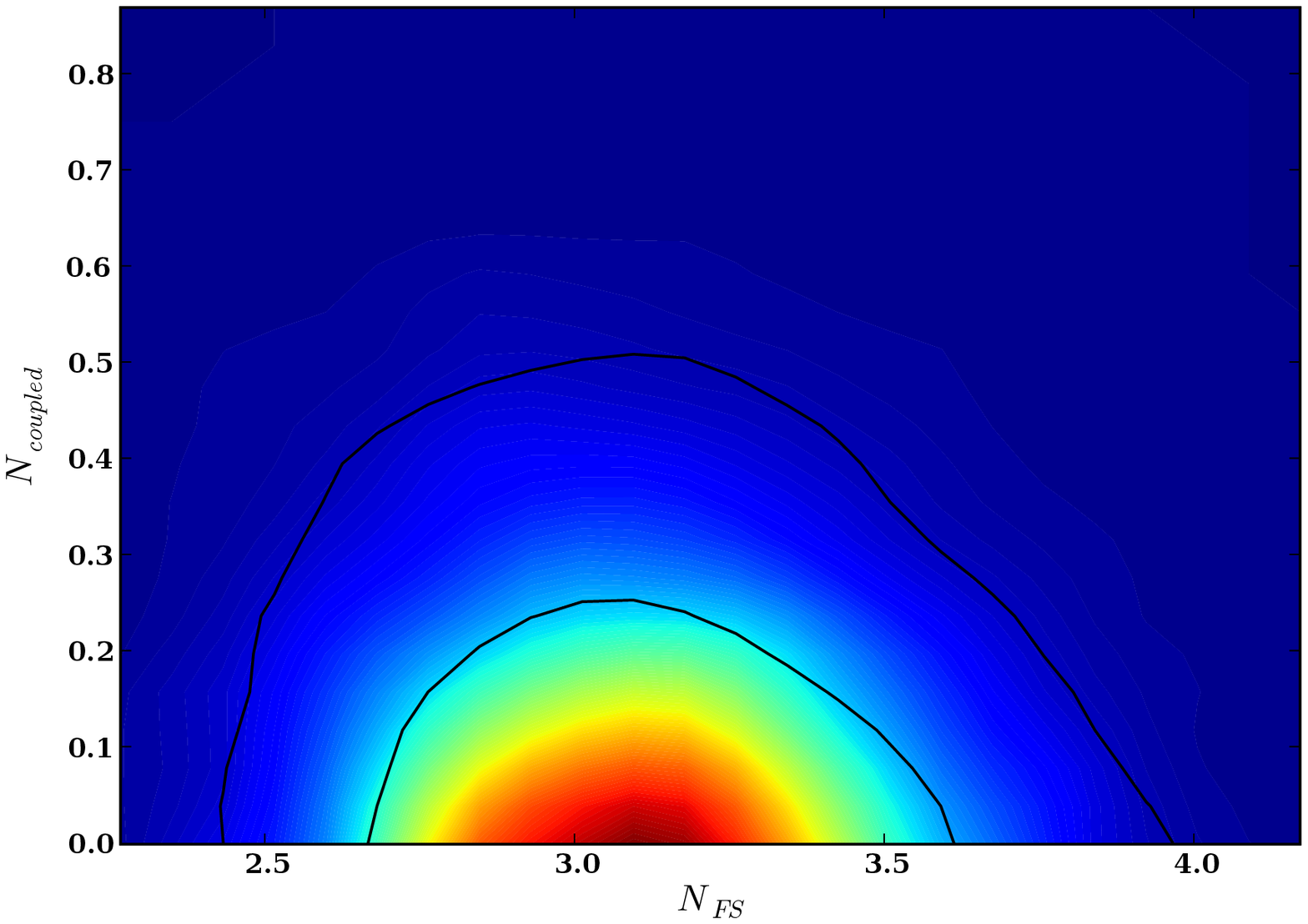}
  \caption{Sensitivity of Planck plus other cosmological data.  The constraints on $N_{FS}$ are not much changed as compared with Planck alone, but the figure shows a factor of two potential improvement in $N_{coupled}$.
  }
  \label{fig:Planck_all_scanNu_coupled_2d}
\end{figure*}

\begin{figure*}[htbp]
  \centering
  \includegraphics[width=0.77\textwidth]{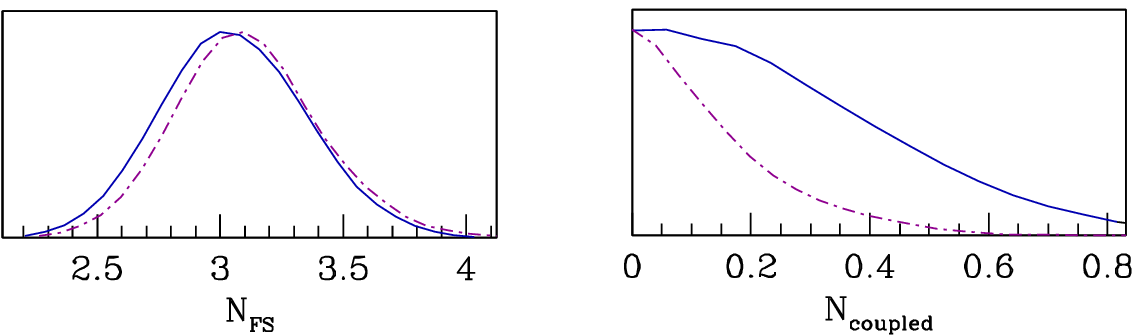}
  \caption{Sensitivity of Planck plus other cosmological data (dashed-dotted)
   vs. Planck only (solid). 1-d marginalized probabilities.
 }
  \label{fig:Planck_all_scanNu_coupled}
\end{figure*}

Lastly, we briefly consider what might be gained by combining Planck
with the other cosmological data. For that, we run a combined fit to
Planck, LRG and Main samples of SDSS, 2dF, the Type Ia supernova
data and the
Hubble constant measurements from the HST.

The result of the fit is shown in
Fig.~\ref{fig:Planck_all_scanNu_coupled_2d}. The sensitivity to the
number of freely streaming neutrinos is unaffected by the addition of
the other data, while the sensitivity to coupled neutrinos improves by
a factor of two. This can also be seen from the one-dimensional
marginalized plots in Fig.~\ref{fig:Planck_all_scanNu_coupled}.

One should, of course, not overinterpret this result. It should be
kept in mind that the mock data for
Planck was generated from the best-fit point of WMAP3 and we did not
explore the dependence of the fit on this choice. Also, by the time
Planck's data is available, the other experiments could be
updated. Nevertheless, the general lesson is that while most of the
precision will come from Planck, the addition of the other data may
lead to further tightening of the bounds.

\section{Discussion of the results and further numerical investigations.}

The results of the previous Section indicate that Planck
should constrain the numbers of coupled (and freely streaming)
neutrino species much more accurately than what is possible today.
In this Section, we make this statement more robust by investigating
the following points:
\begin{itemize}
\item The sensitivity is an indication that not all of the effects of
  neutrino coupling could be compensated and that the residuals should
  be within the sensitivity reach of Planck's instruments. We recall
  that in the last Section, we varied these parameters: $N_{FS}$,
  $N_{coupled}$, $\omega_b$, $\omega_c$, $\theta$, $\tau$, $n_s$,
  $n_{\rm run}$, $A_s$, $Y_{\rm He}$. We have omitted other
  cosmological parameters such as the curvature $\Omega_{curv}$, or
  the dark energy equation of state $w$. If varying these additional
  parameters introduced additional ways of compensating the effects of
  neutrino coupling, the bounds of the previous Section could be
  weakened. We need to show that this is not the case.
\item We also made assumptions about the performance of Planck's
  detector. We need to check the robustness of our results with
  respect to changing the characteristics of the detectors, such as
  the angular resolution and the detector noise levels. Put another
  way, we need to establish which multipoles $l$ are crucial for this
  measurement.
\end{itemize}
We will investigate these points using the Fisher matrix technique.

In addition to these practical issues, we will also investigate the
issue of principle: what is the nature of Planck's sensitivity to
neutrino coupling? We do that by examining which effects of coupling
cannot be compensated by adjusting the cosmological parameters.
\subsection{The role of other cosmological parameters}
\label{sect:Fisher}

As mentioned before, we will investigate the role of other
cosmological parameters using the Fisher matrix technique. The idea
is very simple: suppose that the likelihood function around the
best-fit point is approximately gaussian. In this case, one can fix
the parameters of the gaussian (in the case of an $N$ dimensional
parameter space, an $N\times N$ symmetric matrix) by evaluating the
likelihood function at $O(20)$ points, rather than at $10^5$ points,
as in the case of mapping out the parameter space with MCMC.

The main reason for using this approximate method is speed
\cite{Bowenetal}. Obviously, the Fisher matrix method has its
limitations, as for example, was recently discussed in
\cite{Perotto:2006rj}.  Still we believe that our usage of the Fisher
matrix approximation -- not to obtain the exact bounds but to
investigate qualitative issues outlined above -- is appropriate.

By expanding the log of the likelihood function to the second order
in cosmological parameters $s_i$ around the maximum, one finds the
standard expression:
 \be
  F_{ij} = \sum_l\sum_{X,Y}\frac{\partial C_l^X}{\partial s_i}
 \mbox{Cov}^{-1}(\hat{C}_l^X,\hat{C}_l^Y)\frac{\partial
   C_l^Y}{\partial s_j}.
\label{eq:ourFisher}
\end{equation}
Here $\mbox{Cov}^{-1}$ is the inverse of the covariance matrix (to be
defined shortly), and the $C_l^X$ are the power spectra in the
temperature and polarization channels.  We restrict ourselves to the
temperature, $T$, $E$ polarized, and cross temperature-polarization
$C$, power spectra. They are the only ones relevant for scalar
perturbations.

The elements of the covariance matrix
\cite{Seljak:1996ti,Zaldarriaga:1996xe,Eisenstein:1998hr}, which give the errors of the
corresponding measurements, are: 
 \begin{eqnarray}
   \label{eq:covmatrixTT}
   (\Cov_\ell)_{TT}&=&{2\over(2\ell+1)\fsky}(C_{T\ell}+w_T^{-1}\beam^{-2})^2, \\
   \label{eq:covmatrixEE}
(\Cov_\ell)_{EE}&=&{2\over(2\ell+1)\fsky}(C_{E\ell}+w_P^{-1}\beam^{-2})^2, \\
\label{eq:covmatrixCC}
(\Cov_\ell)_{CC}&=&{1\over(2\ell+1)\fsky}\left[C_{C\ell}^2
    +(C_{T\ell}+w_T^{-1}\beam^{-2})
    \right.\nonumber\\&&\left.
    \times(C_{E\ell}+w_P^{-1}\beam^{-2})\right],\\
    \label{eq:covmatrixTE}
(\Cov_\ell)_{TE}&=&{2\over(2\ell+1)\fsky}C_{C\ell}^2, \\
\label{eq:covmatrixTC}
(\Cov_\ell)_{TC}&=&{2\over(2\ell+1)\fsky}C_{C\ell}
        (C_{T\ell}+w_T^{-1}\beam^{-2}), \\
        \label{eq:covmatrixEC}
(\Cov_\ell)_{EC}&=&{2\over(2\ell+1)\fsky}C_{C\ell}
        (C_{E\ell}+w_P^{-1}\beam^{-2}).
 \end{eqnarray}
In these equations, $f_{sky}$ is the sky coverage, $w_T^{-1}$ and
$w_P^{-1}$ specify the detector noise for temperature and
polarization respectively, and $B_\ell^{-2}=
 e^{l(l+1)\theta_b^2/(8\ln 2)}$ is the beam smearing window function.
Here $\theta_{b}$ is the full-width,
 half-maximum of the beam in radians, $w_T$ and $w_P$ are the
 quantities characterizing the detector noise level for temperature
 and polarization, respectively.  For Planck, we take $\theta_b = 7'$, $w_T^{-1/2} = 40 \mu\mbox{K}'$, and $w_P^{-1/2} = 56\mu\mbox{K}'$.

More precisely, the Fisher information matrix $F_{ij}$ is defined
through derivatives of the likelihood function ${\cal L}(x,p)$
for data~$x$ and model parameters~$p$ as
\begin{equation}
F_{ij}\equiv \left\langle -\frac{\partial^2 \ln {\cal L}(x,p)}{\partial p^i\partial p^j}
                 \right\rangle_{x}.
\label{FishM_def}
\end{equation}
The right-hand side of Eq.~(\ref{FishM_def}) is averaged over the
data~$x$, weighted by the probability ${\cal L}(x,p)$ of their observation in
the fiducial model. Given this definition, the Cram\'er-Rao inequality
states that the r.m.s. of the best-estimator for a parameter~$p^i$
cannot be less than $\sqrt{({F^{-1}})^{ii}}$, as discussed, {\it
  e.g.}, in \cite{Eisenstein:1998hr}.

Of course, to perform the average in general requires mapping out
the likelihood function over the parameter space and we are back to
the problem of computing the likelihood at $O(10^5)$ points. We do not perform such mapping
here. Instead, we simply assume the likelihood is close to gaussian,
compute the matrix in Eq.~(\ref{eq:ourFisher}), and use
$\sqrt{({F^{-1}})^{ii}}$ as an estimate of a 1$\sigma$ error on the
parameter $s_i$.

We consider the following set of cosmological parameters
$(\omega_m/\omega_r, \omega_b,\Omega_{\rm
  de}, N_\nu^{eff}, N_{coupled}, n_s,\alpha_s$ $ A_s, Y_{\rm He}, \tau, \omega_{\rm \nu}, \Omega_{\rm c},
w_{\rm de})$, replacing $N_{FS}$ with $N_\nu^{eff} = N_{FS}+N_{coupled}$ in our list of parameters.  The
first 10 parameters span the parameter space of our MCMC analysis of
the last Section, while the last three are the new additions. We
calculate the derivatives in Eq.~(\ref{eq:ourFisher}) by symmetric
finite differences about the best-fit cosmological parameters from
the WMAP year three data. We compute the resulting 1$\sigma$
marginalized error on $N_{coupled}$ and $N_\nu^{eff}$ with all 13 parameters and then
drop the last three parameters in turn. We also consider the effect
of fixing the helium abundance $Y_{\rm He}$.

\begin{table*}[hbt]
\begin{tabular}{|l||c|c|c|c|c|} \hline
 & Vary all  & Fix & Fix $\omega_\nu =0 $, & Fix $\omega_\nu =0 $, $\Omega_{curv}=0$,  &
 Fix $\omega_\nu =0 $,
$\Omega_{curv}=0$,  \\

& 13 params.& $\omega_\nu =0 $ & $\Omega_{curv}=0$ & $w_{\rm de}=-1$ &

$w_{\rm de}=-1$, $Y_{\rm He}=0.24$\\
\hline Error on $N_{coupled}$ &  0.31 & 0.31& 0.31 & 0.29  & 0.28\\
\hline Error on $N_\nu^{eff}$ & 0.38 & 0.38 & 0.38 & 0.35 & 0.32 \\
\hline
\end{tabular}
\caption{The Fisher matrix estimate of the 1$\sigma$ error on $N_{coupled}$ and $N_\nu^{eff}$. The effects of fixing neutrino mass, curvature,
  dark energy equation of state, and the Helium fraction are considered.}
\label{table:3extraparams}
\end{table*}

The results are shown in Table~\ref{table:3extraparams}. We see
that keeping the curvature and the neutrino mass fixed to zero is
completely justified. Moreover, the effect of varying the dark
energy equation of state is also quite small ($\sim 7$\% for
$N_{coupled}$ and $\sim 9\%$ for $N_\nu^{eff}$), as is the effect of
varying $Y_{\rm He}$ ($\sim 4$\% for $N_{coupled}$ and $\sim 9\%$
for $N_\nu^{eff}$).  Thus, practically speaking, we are justified in
our choice of the 10 cosmological parameters used in our MCMC
calculations, as the other three parameters do not change our
qualitative conclusions about Planck's sensitivity to the number
of coupled neutrinos.  In addition, though the Fisher analysis
shows that the effects of $Y_{\rm He}$ on the errors of
$N_{coupled}$ and $N_\nu^{eff}$ are fairly minimal, we include
$Y_{\rm He}$ as a parameter in our MCMC with priors consistent
with astrophysical observations as described in
Sect.~\ref{sect:WMAPplusEBCS}.

\subsection{The compensation mechanism and the nature of Planck's sensitivity}
\label{sect:physicsofsensitivity}

We now wish to establish the nature of Planck's sensitivity to
neutrino self-coupling. By this we mean showing which effects of
self-coupling cannot be entirely compensated by adjusting the
cosmological parameters and how the residual differences compare to
the corresponding errors. The latter, as we saw in the previous
subsection, are set by a combination of the cosmic variance and the
resolution/sensitivity of the apparatus. Equipped with our numerical
results, we are now able to explicitly see the compensation
mechanism in action.

For this, we turn to Table~\ref{table:bestfitPlanckonly}. The
best-fit parameters listed in the Table are precisely those which
compensate the effects of neutrino coupling most efficiently within
each scenario. Indeed, we recall that while the mock data is always
generated under the assumptions of three freely streaming neutrinos,
for each of the scenarios listed in the Table, the MCMC code
attempts to find the best fit possible within the framework of that
particular scenario. The code varies the values of the cosmological
parameters until the closest agreement is found.

\begin{figure}[htbp]
  \centering
  \includegraphics[width=0.49\textwidth]{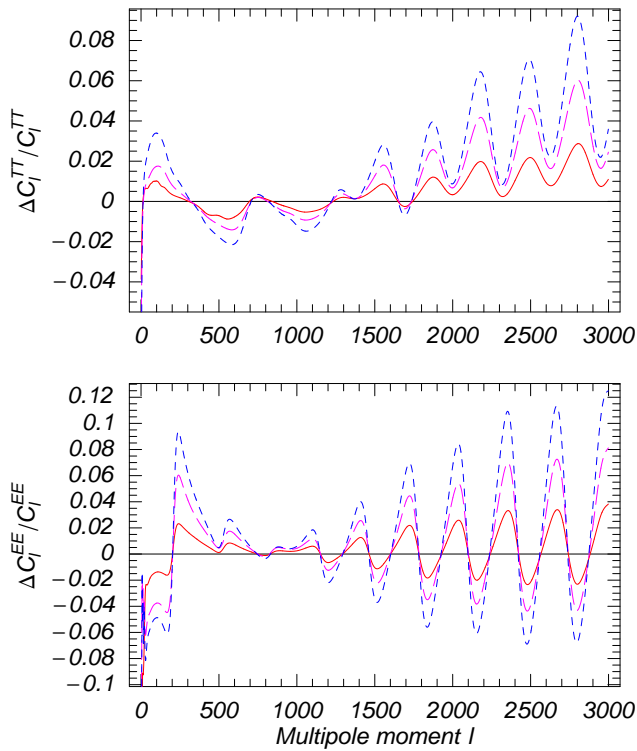}
  \caption{The relative residual differences in temperature and
  polarization power spectra for 1, 2, and 3 coupled neutrinos
  (in order of deviation from zero) that remain after the cosmological
  parameters are adjusted in each of the scenarios
  (as shown in Table~\protect{\ref{table:bestfitPlanckonly}}).}
  \label{fig:clsrelative}
\end{figure}

The compensation turns out to be very efficient. One way to
illustrate it is to plot the relative differences of the temperature
and polarization power spectra in the scenarios of coupled vs.
freely streaming neutrinos that remain after the cosmological
parameters are adjusted. This is done in Fig.~\ref{fig:clsrelative},
where we show the situations for  1, 2, and 3 coupled neutrinos. We
can see that, unlike the differences shown in
Fig.~\ref{fig:illustrate_coupling} which were in tens of percent,
these residual differences are only at the level of 1-3\%.

We further observe that below the multipole number of $l\sim 1300$
the residuals have no clear structure, while above they develop a
shape of periodic oscillations, as a consequence of
the phase shift, Eq.~(\ref{eq:phaseshift}).

In terms of the sensitivity of Planck, the physically relevant
quantities are not the relative differences in the $C_l$'s just
shown, but the ratios of the differences to the corresponding
errors. The errors are given by the elements of the covariance
matrix, Eqs.~(\ref{eq:covmatrixTT}-\ref{eq:covmatrixEC}). We can
then consider at each $\ell$ the quantities
 \be
  (\delta\chi_\ell)^{XX} \equiv \frac{\Delta
  C_l^X}{\sqrt{(\Cov_\ell)_{XX}}},
   \label{eq:deltaovererror}
\end{equation}
where $X$ runs over $T$, $E$, and $C$, and $\Delta C_l^X$ are again
the differences between the power spectra of the
  reference model ($N_{FS}=N_\nu^{eff}=3$) and the best-fit power
  spectra in a scenario with self-coupled neutrinos.

\begin{figure}[tbp]
  \centering
  \includegraphics[width=0.49\textwidth]{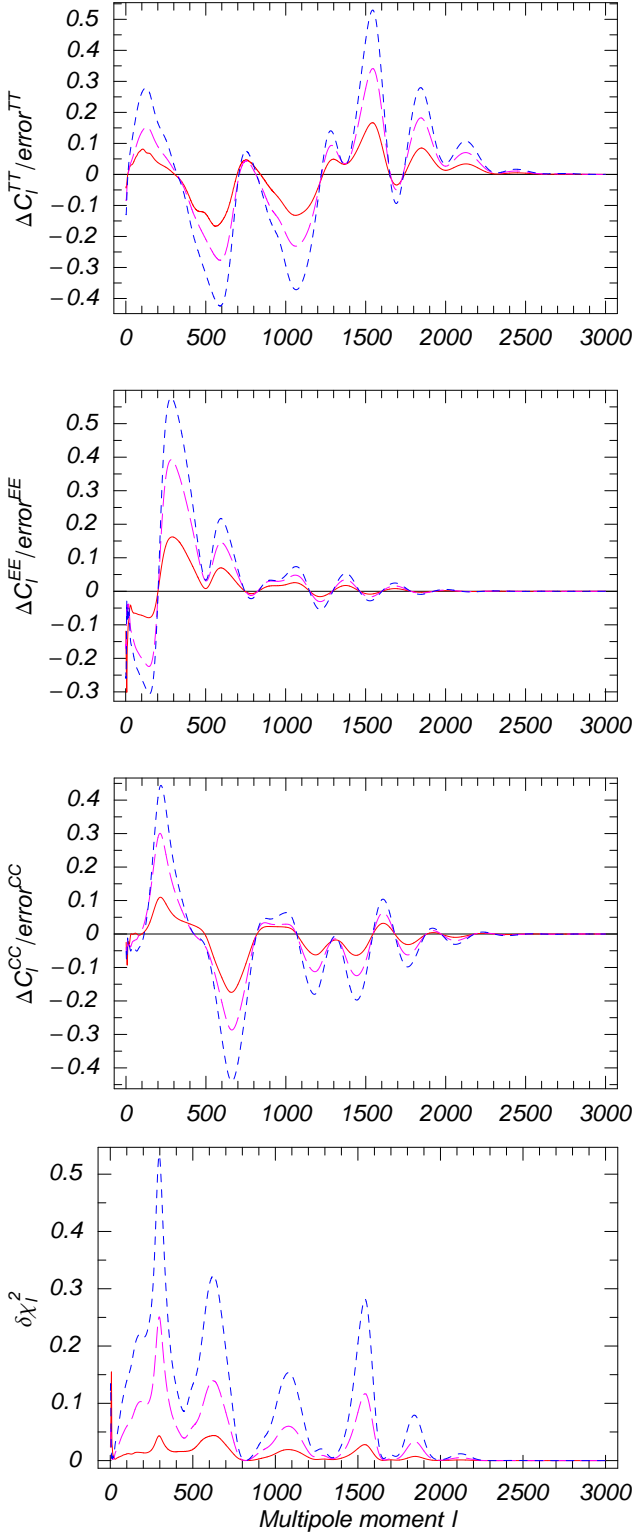}
  \caption{The quantities $(\delta\chi_\ell)^{X}$ and $(\delta\chi_\ell^2)$
  defined in Eqs.~\protect{\ref{eq:deltaovererror}} and \protect{\ref{eq:deltachi2}}.
  They illustrate the ratios of the residual discrepancies at each $\ell$
  to the corresponding errors, as explained in the text. The residual
  discrepancies at $\ell\lesssim 1300$ play a key role.}
  \label{fig:deltaclsovererror}
\end{figure}

The quantities $(\delta\chi_\ell)^{X}$ are plotted in
Fig.~\ref{fig:deltaclsovererror} (the top three panels). In the last
panel of the figure, we plot the quantity
 \be
  (\delta\chi_\ell^2) \equiv \sum_{X,Y=\{T,E,C\}}(\Delta
  C_\ell^X)(\Cov_\ell)_{XY}^{-1}(\Delta C_\ell^Y),
   \label{eq:deltachi2}
\end{equation}
which is nothing but the contribution of a given multipole $\ell$ to
the $\chi^2$.  These Figures are very instructive, for they show that
statistically significant deviations occurs for $\ell\lesssim 1300$,
and have no clear structure.  The higher multipoles which exhibit
clear effects of the phase shift in Fig.~\ref{fig:clsrelative} turn
out to be of relatively minor importance.

\begin{figure}[tbp]
  \centering
  \includegraphics[width=0.44\textwidth]{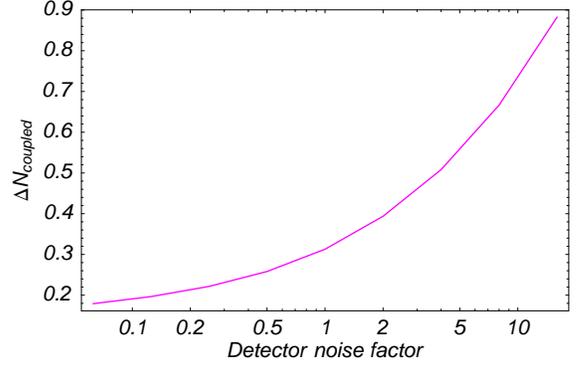}
  \caption{$1\sigma$ sensitivity of Planck (calculated with a Fisher
    matrix) to $\Delta N_{coupled}$ as a function of the detector
    noise.  A noise factor of one corresponds to the sensitivity
    chosen in the main analysis ($w_T^{-1/2} = 40 \mu K'$, 
    $w_P^{-1/2} = 56 \mu K'$).}
  \label{fig:varynoise}
\end{figure}

\begin{figure}[tbp]
  \centering
  \includegraphics[width=0.44\textwidth]{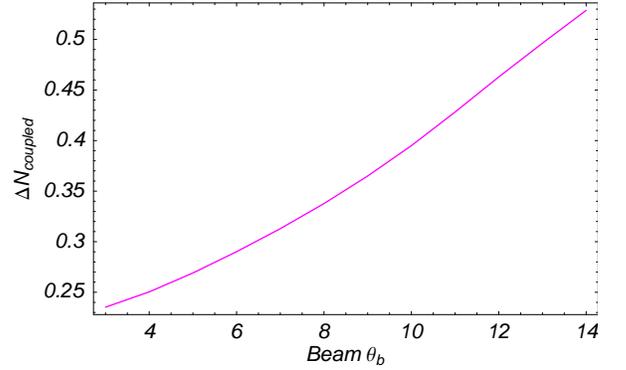}
  \caption{$1\sigma$ sensitivity of Planck (calculated with a Fisher
    matrix) to $\Delta N_{coupled}$ as a function of the detector 
    resolution.  The resolution chosen in the main analysis is 
    $ 7'$.}
  \label{fig:varyresolution}
\end{figure}

\begin{figure}[tbp]
  \centering
  \includegraphics[width=0.44\textwidth]{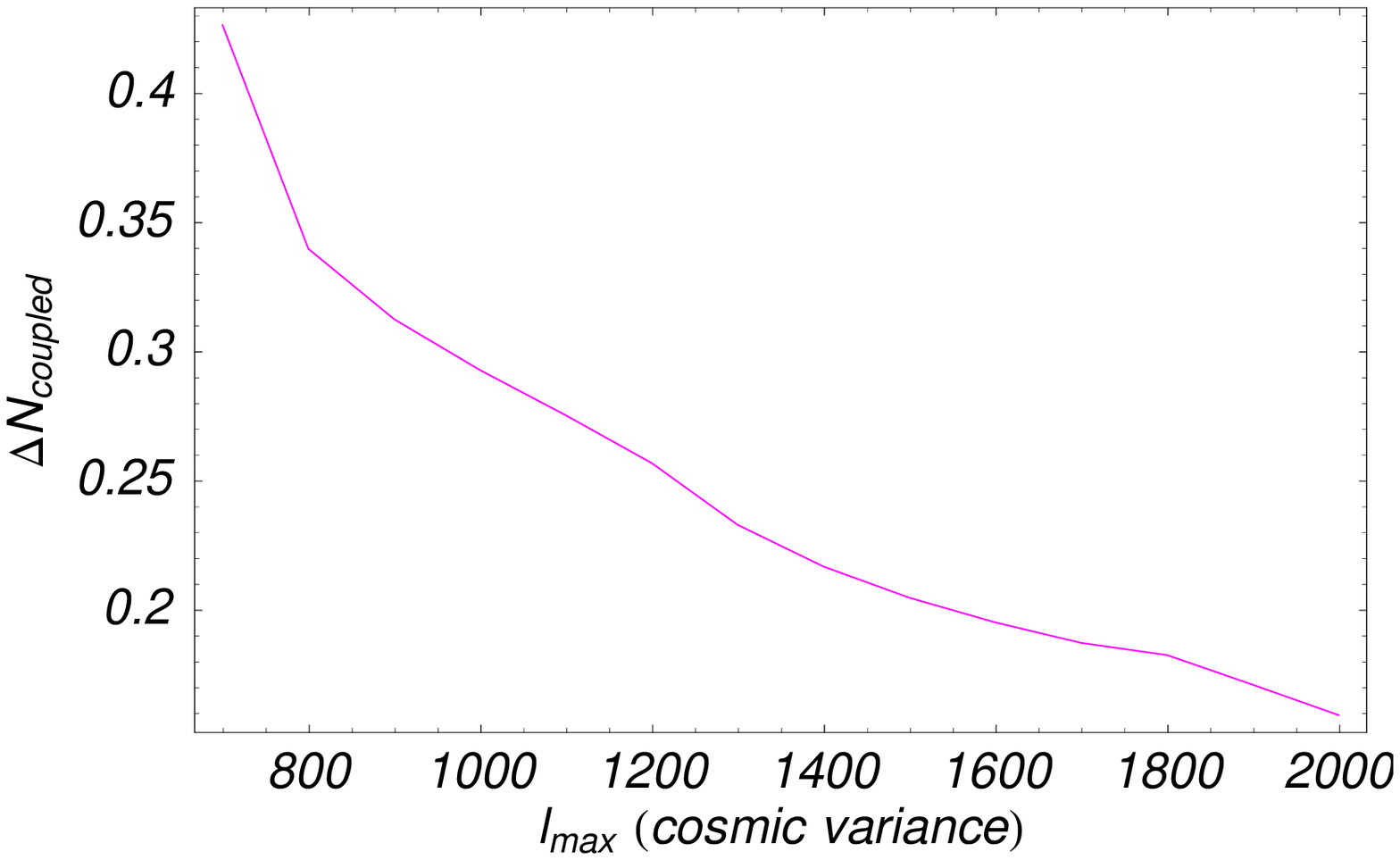}
  \caption{$1\sigma$ sensitivity of a hypothetical CMB experiment 
    to $\Delta N_{coupled}$ as a function
    of the maximum multipole, $l_{max}$, measured to within cosmic
    variance. Multipoles with $l>l_{max}$ are discarded.}
  \label{fig:cosmovar}
\end{figure}

This is further seen when we consider the effects of the expected
detector resolution and noise on the $1\sigma$ errors in
$N_{coupled}$.  In
Figs.~\ref{fig:varynoise},~\ref{fig:varyresolution}, we show the
change in sensitivity as a function of detector characteristics,
obtained a Fisher analysis. The rate at which increased resolution and
lower noise improve the sensitivity to $\Delta N_{coupled}$ is rather
modest.
Decreasing the noise by an order of magnitude, for example, improves
the sensitivity by less than a factor of two.
In addition, we consider the sensitivity of
a hypothetical experiment which measures all multipoles with $\ell <
\ell_{max}$ within cosmic variance; we see in Fig.~\ref{fig:cosmovar}
that sensitivity to multipoles $\ell_{max} \lesssim 1000$ is
sufficient to measure $\Delta N_{coupled} = 0.3$, and implies that
Planck's high sensitivity to neutrino free-streaming does not depend
on the high $\ell$ multipoles.  It also implies that our qualitative
predictions for Planck have a high degree of robustness, as the
sensitivity does not depend strongly on the high $\ell$ properties of
the detector.

One characteristic of Planck that proves crucial is the
simultaneous measurement of both temperature and polarization. This
makes it possible for Planck to constrain $N_{coupled}$ down to 0.5
at 1$\sigma$ level even if $l_{\rm max}$ is only 800. In comparison,
we saw that WMAP3 places virtually no constraint on the number of
self-coupled neutrinos. The difference is the relatively poor
quality of the polarization measurements at WMAP, as well as the
temperature measurements for $500\lesssim l \lesssim 800$,
where Planck is cosmic variance limited.

\section{Constraints on neutrino interactions}
\label{sect:discussconstraints}

Since we have shown that even a single interacting neutrino or a
single extra neutrino will be observed or excluded to high
precision with Planck, we consider the implications for models of
neutrino-scalar interactions, including neutrino dark energy.
Before turning to the specific models, we first review general
constraints on the couplings $g$ from Majorana neutrino-scalar
interactions which will be possible with Planck data.

The loss of neutrino free-streaming and production of additional
relativistic degrees of freedom as a result of neutrino-scalar
coupling have been considered in Ref.~\cite{Chacko:2003dt}. We
summarize and reformulate the constraints here; a more complete treatment is left to the appendix.

For a Majorana mass term, the relevant interactions $\nu \nu
\leftrightarrow \phi\phi$, $\nu \nu \leftrightarrow \nu \nu$, and
(if $m_\phi > 2 m_\nu$) $\nu \nu \leftrightarrow \phi$.  The last
interaction is often most constraining on the coupling $g$, and its
rate is 
\be 
\Gamma(\nu \nu \leftrightarrow \phi) \sim \frac{g^2}{16\pi}\frac{m_\phi}{T}m_\phi 
\label{two2one} 
\ee 
for $m_\phi<T$.  This rate can also be described from on resonant s-channel scattering; see the appendix for details.  The
$m_\phi/T$ factor accounts for the boost from the rest frame of
the scalar.  This rate increases as the temperature decreases, and will cause $\phi$ to decay to neutrinos, increasing
$N_\nu^{eff}$, once $m_\phi$ drops below $T$, provided $g
\gtrsim 10^{-13}(T_{rec}/m_\phi)$.

If, on the other hand, $m_\phi$ remains relativistic through
recombination, this same process may tightly couple the neutrino to
the scalar, removing neutrino free-streaming.  The $2 \rightarrow 1$
process can only occur in a small region of phase space which
depends on the angle between the incoming neutrinos, $\theta \sim
m_\phi/T$.  In order to isotropize the neutrino momentum, therefore,
this process must occur $N= (T/m_\phi)^2$ times.  We require then
$\Gamma(\nu \nu \leftrightarrow \phi) > N H(T_{rec})$
\cite{Chacko:2003dt}.  The equilibration occurs before recombination
if $g \gtrsim 10^{-13}$ for $m_\phi \sim T_{rec}$, and
becoming less restrictive with dropping $m_\phi$ by the factor
$(T_{rec}/m_\phi)^2$.

The exception is if scattering of $\phi$'s through a $\phi^4$
interaction may be fast enough to isotropize the scalar momentum in
between decays and inverse decays.  Then the constraint is
transmuted to $\Gamma(\phi \phi \leftrightarrow \phi \phi) >
\Gamma(\nu \nu \leftrightarrow \phi) > H(T_{rec})$, which
implies that if $\lambda_\phi^2 T > T^2/M_{pl}$, where $\lambda_\phi$ is
the coupling of the $\phi^4$ interaction, and if $g^2 m_\phi^2/T > T^2/M_{pl}$, then the neutrinos will be strongly coupled with the scalars. The equilibration then
occurs before recombination if 
\be
g \gtrsim 10^{-14} (T_{rec}/m_\phi)
\ee
and 
\be
\lambda_\phi \gtrsim 10^{-14}.
\ee

$2 \leftrightarrow 2$ interactions offer complementary constraints
to the decay-inverse decay processes.  If both neutrinos and
scalars are relativistic at decoupling, the rate for $\nu \nu
\leftrightarrow \nu \nu$ is given by
 \be
  \Gamma(\nu \nu \leftrightarrow \nu \nu) \sim \frac{g^4}{16 \pi} T,
 \label{two2two}
 \ee
and is competitive with $\nu \nu \leftrightarrow \phi \phi$,
so that if $g \gtrsim 10^{-7}$ neutrinos and scalars will be
tightly coupled.  If $m_\phi \gtrsim T_{rec}$, however, the
constraints on $g$ are relaxed by $T/m_\phi$, due to propagator suppression from the scalar.

We have assumed everywhere that $m_\nu \ll T_{rec}$. If $g \gtrsim
10^{-5}$ and $m_\nu > T_{rec}, \mbox{ } m_\phi$, the neutrinos will
annihilate to the scalars, producing a ``neutrinoless'' universe
\cite{Beacom:2004yd}; for this process, $N_\nu^{eff}$ is an evolving
function of redshift as the neutrinos annihilate. We do not consider
this scenario further in this paper.

We briefly discuss the constraints from BBN.  The precise upper bound
on the effective number of neutrinos is somewhat controversial and
dependent on the particular set of constraints from data utilized, but
is on the order of $N_\nu^{eff} \lesssim 4.5$ \cite{Cyburt:2004yc}.
Any additional scalar degrees of freedom brought into equilibrium
before BBN will contribute $\delta N_\nu^{eff} = 4/7$.  One scalar
alone brought into thermal equilibrium before BBN, then, will not
upset the current bounds.  If the theory contains multiple scalars,
however, there will be constraints already from BBN on the coupling
$g$ from $\nu \nu \leftrightarrow \phi$ and $\nu
\nu \leftrightarrow \phi \phi$ coming from demanding that these processes not recouple until $T< T_{BBN}$. 

If these processes are brought into equilibrium after the neutrino
processes decouple from the heat bath at $T \sim 1 \mbox{ MeV}$, they
may still affect the value of $N_\nu^{eff}$ measured by the CMB
\cite{Chacko:2003dt,Chacko:2004cz}. For example, it may be possible
that new particle states ({\it e.g.}, scalars, sterile neutrinos) are
populated by recoupling after BBN and then annihilate when the
temperature drops below their masses (but still before the CMB
decoupling). In this scenario, while the annihilation step occurs at
constant entropy, the recoupling does not. In the approximation when
the former occurs at constant energy, the value of $N_\nu^{eff}$
measured by the CMB is given by Eq.~(8) of \cite{Chacko:2004cz}. In
general, for a given model, accurate analysis of recoupling may be
required.

We summarize the future constraints from Planck given no deviation
from standard relativistic energy density and neutrino
free-streaming in Table~\ref{table:CMBconstraints}.

\begin{table*}[t]
\begin{tabular}{|c|c|c|c|}\hline
 process & $m_\phi$  & constraint & Effect on CMB \\ \hline
$\nu \nu \leftrightarrow \phi$ & $T_{rec} > m_\phi$ & $g \lesssim 10^{-13} \left(1\mbox{ eV}/m_\phi\right)^2$ & Remove free-streaming \\
& $T_{rec} < m_\phi$ & $g \lesssim 10^{-13}
\left(m_\phi/1\mbox{ eV}\right)^{1/2}$ & Increase
$N_\nu^{eff}$ \\ \hline
$\nu \nu \leftrightarrow \nu \nu$ & $T_{rec} > m_\phi$ & $g \lesssim 10^{-7}$ & Remove free-streaming \\
 & $T_{rec} < m_\phi$ & $g \lesssim 10^{-7} \left(m_\phi/1\mbox{ eV}\right)$ & Remove free-streaming \\  \hline
$\nu \nu \rightarrow \phi \phi$ & $ m_\phi < T_{rec} <
m_\nu$ & $ g \lesssim 10^{-5}$ & Increase $N_\nu^{eff}$ \\
\hline
\end{tabular}
\caption{Constraints on the effective coupling $g$ of a Majorana
neutrino with a scalar, if the Planck satellite observes no extra,
non-standard relativistic degrees of freedom or strong
neutrino-scalar interactions.}
 \label{table:CMBconstraints}
\end{table*}

\section{Some implications for models}
\label{sect:implicationsformodels}

\subsection{Models of Neutrino Dark Energy}

Models of neutrino dark energy are typically characterized by a coupling
between a singlet neutrino and a light scalar field of the form
given in Eq.~\ref{dynamicalseesaw}.  An important difference between the standard see-saw and the MaVaN scenario is that $m_N$ is a continuously evolving parameter, since the vev of the singlet scalar, $\langle \phi \rangle$, varies as the universe cools.

The evolution of $\langle \phi \rangle$ results from finite temperature effects in the scalar potential; the background neutrino energy density acts as a source for the scalar field, and may displace $\langle \phi \rangle$ from the value dictated by its zero-temperature potential, $V_0(\phi)$:
\begin{eqnarray}
V(\phi) & = & V_0(\phi) +\rho_\nu(\phi) \nonumber \\
& \simeq & \frac{1}{2} m_\phi^2 \phi^2 + \frac{7 \pi^2}{120} T^4 +
\frac{m_\nu^2(\phi) T^2}{24},
\label{potential}
\end{eqnarray}
where $m_\nu$ is the light mass eigenstate resulting when the sterile neutrino is integrated out:
\begin{eqnarray}
m_\nu & = & \frac{m_D^2}{m_N} \nonumber \\
 & = & \frac{m_D^2}{\lambda \langle \phi \rangle}.
\end{eqnarray}
This is a self-consistent formalism provided that the sterile neutrino remains thermally unpopulated (and hence effectively integrated out) in the early universe; it was shown in \cite{Weiner:2005ac} how this can be done with Planck suppressed operators between the scalar and electrons.
In addition, in a theory with the full flavor structure of three generations, $m_\nu$ must correspond to the lightest mass eigenstate, which is
still relativistic in the universe today, so that the neutrino dark energy does not clump and become unstable \cite{Afshordi:2005ym}.

As was shown in Sect.~\ref{sect:neutrinomodels}, the mass eigenstate couples to the scalar through the Majoran interaction, ${\cal L} = g \phi \nu \nu$, where $g \sim \lambda (m_D/m_N)^2$.  Minimizing eqn.~\ref{potential},
\begin{eqnarray}
m_N & = & \lambda \langle \phi \rangle \nonumber \\
& = & m_D \sqrt{\frac{\lambda T}{m_\phi}}
\label{sterilemass}
\end{eqnarray}
Hence, we have a {\it temperature dependent} coupling $g$.

How can MaVaNs be detected with Planck?  Planck will be sensitive to couplings g, through $\nu \nu \leftrightarrow \nu \nu$, of size $g \gtrsim 10^{-7}$; these couplings may be small enough to lead to possible detection or interesting constraints on MaVaNs with Planck.  We conclude, using $m_N$ from eqn.~\ref{sterilemass}
\be
g = \frac{m_\phi}{T},
\ee
implying that the neutrinos will form a tightly coupled fluid down to recombination temperatures, $T \sim 0.1 \mbox{ eV}$, unless
\be
m_\phi  \lesssim 10^{-8} \mbox{ eV}.
\label{phimassconstraint}
\ee
For scalars in these mass ranges, Planck will be able to detect MaVaNs.

What are natural parameters and typical scalar masses for the theory?   We consider a generic class of MaVaN models laid out in \cite{Fardon:2003eh,Fardon:2005wc}.  All energy scales in MaVaN models are characterized by the dark energy scale, $\Lambda \sim 10^{-2.5}\mbox{ eV}$.  All parameters in the model, including $m_D$ and UV cutoff of the neutrino sector, are characterized by meV scale physics, in contrast to the traditional see-saw, which is characterized by a sterile neutrino at the GUT scale.  Radiative
corrections will drive the scalar mass to the cutoff of
the theory, implying that without an unnatural fine-tuning, the
scalar mass should not be much lighter than the cutoff of the theory
at a meV.  A superpotential which radiatively controls the scalar mass is simply the see-saw, eqn.~\ref{dynamicalseesaw}, with all the fields promoted to superfields:
\be W = \lambda \phi NN + y H L N. \ee
Assuming $\langle| N | \rangle = 0$, the superpotential generates no tree-level mass term for $\phi$.  A mass is generated radiatively for $\phi$, however, through a 1-loop diagram with an active and sterile sneutrino in the loop
so that
\begin{equation}
  m_\phi^2 \sim \frac{\lambda^2 m_D^2}{16 \pi^2}\log \Lambda^2/m_D^2
  \label{eq:1loopcorr}
\end{equation}
where $\Lambda$ is the UV cutoff of the neutrino sector, typically around an eV.  We can see that loop corrections will generically generate a mass for the scalar, $m_\phi \sim \lambda m_D$.

Besides the loop corrections of Eq.~(\ref{eq:1loopcorr}) one has to
keep in mind corrections mediated by gravity. The natural size of
gravity-mediated corrections to the scalar mass from supersymmetry
broken at the TeV scale is $\sim F/M_{pl}\sim 10^{-3}$ eV, or some
five orders of magnitude above the bound in
Eq.~(\ref{phimassconstraint}).

$\lambda$ and $m_D$ are themselves constrained by the requirement that MaVaNs produce dark energy.  In particular, it was shown \cite{Fardon:2005wc} that a mass term appears for $|N|$ through radiative corrections of size $-\lambda^2 m_D^2$, as it does for $|\phi|$, but with the opposite sign.  This implies a false minimum at $\langle | N | \rangle = 0$, with a true minimum at $\langle | N | \rangle \sim m_D/\lambda$ ; the energy difference between the true minimum of the potential  and $\langle | N | \rangle = 0$ is the dark energy density,
\be
\rho_\Lambda \sim m_D^4/\lambda^2 \sim 10^{-10} \mbox{ eV}^4
\ee
and, using $m_\phi \sim \lambda m_D$ from naturalness, we find that the only underlying free parameter is the coupling $\lambda$: $m_\phi \sim \lambda m_D \sim \lambda^{3/2} 10^{-2.5} \mbox{ eV}$.
In terms of the Lagrangian parameter $\lambda$, then, Planck will be able to constrain the coupling to be
\be
\lambda  \lesssim   10^{-5.5}.
\ee
Since $\lambda$ is a parameter typically envisioned to be not much smaller than 1, the reach of Planck is considerable.  Of course, these constraints on Lagrangian parameters depend on naturalness requirements; the direct (more model independent) constraint is expressed in terms of the scalar mass in eqn.~\ref{phimassconstraint}.

The same parameter $\lambda$ enters into matter effects in neutrino oscillation experiments, studied in, e.g., \cite{Kaplan:2004dq,Zurek:2004vd,Barger:2005mn,Cirelli:2005sg,Blennow:2005qj}.  There it was shown that the shift of the neutrino mass due to earth size matter effects is
\be
\Delta M = 1 \mbox{ eV} \left( \frac{\lambda}{10^{-1}}\right)\left( \frac{\alpha_N}{10^{-2}}\right)\left( \frac{\rho_N}{\rho_N^0}\right)\left( \frac{10^{-6} \mbox{ eV}}{m_\phi}\right)^2,
\label{mattereffects}
\ee
where $\alpha_N$ is a Planck suppressed coupling between nucleons and $\phi$, and $\rho_N^0 = 3 \mbox{g/cm}^3$ is the earth energy density.  Neutrino oscillation experiments in earth typically cannot reach sensitivities to mass scales smaller than $10^{-1}-10^{-3} \mbox{ eV}$.   From this, we can can see that Planck and neutrino oscillation experiments are complimentary probes of neutrino dark energy.  Neutrino oscillation experiments are efficient at detecting effects in the low $m_\phi$ region of parameter space; using the naturalness requirement, $m_\phi \sim \lambda m_D$, and the dark energy requirement, $m_D^4/\lambda^2 \sim \rho_\Lambda$,
\be
\Delta M = 1 \mbox{ eV} \left( \frac{\alpha_N}{10^{-2}}\right)\left( \frac{\rho_N}{\rho_N^0}\right)\left( \frac{10^{-7} \mbox{ eV}}{m_\phi}\right)^{4/3}.
\label{mattereffects2}
\ee
Planck, on the other hand, can probe the large $m_\phi$ region according to eqn.~\ref{phimassconstraint}.

\subsection{Late time neutrino masses}

In models of late time neutrino masses \cite{Chacko:2003dt,Chacko:2004cz}, $\phi$ carries a $U(1)$ charge, which has a low scale $f$ of symmetry breaking.  Since the mass eigenstate in the operator of the form $g \phi \nu \nu$ only acquires a mass once the $U(1)$ is broken (exactly as in the Higgs mechanism), the neutrinos only gain their masses at late time.  For a low-energy Majorana mass term, we expect $g \sim m_\nu / f$, as shown in Sect.~\ref{sect:neutrinomodels}.  We conclude then that Planck will be able to probe scales of neutrino mass generation:
\be
f  \lesssim \begin{array}{cc}  1 \mbox{ MeV} & m_\phi < 2 m_\nu  \\
   1 \mbox{ TeV} & m_\phi > 2 m_\nu \end{array},
\label{fconstraint}
\ee
assuming $m_\nu \sim 0.1 \mbox{ eV}$.  Hence Planck may probe TeV scale neutrino mass generation.

\section{Conclusions}

\begin{figure*}[htbp]
  \centering
  \includegraphics[width=0.49\textwidth]{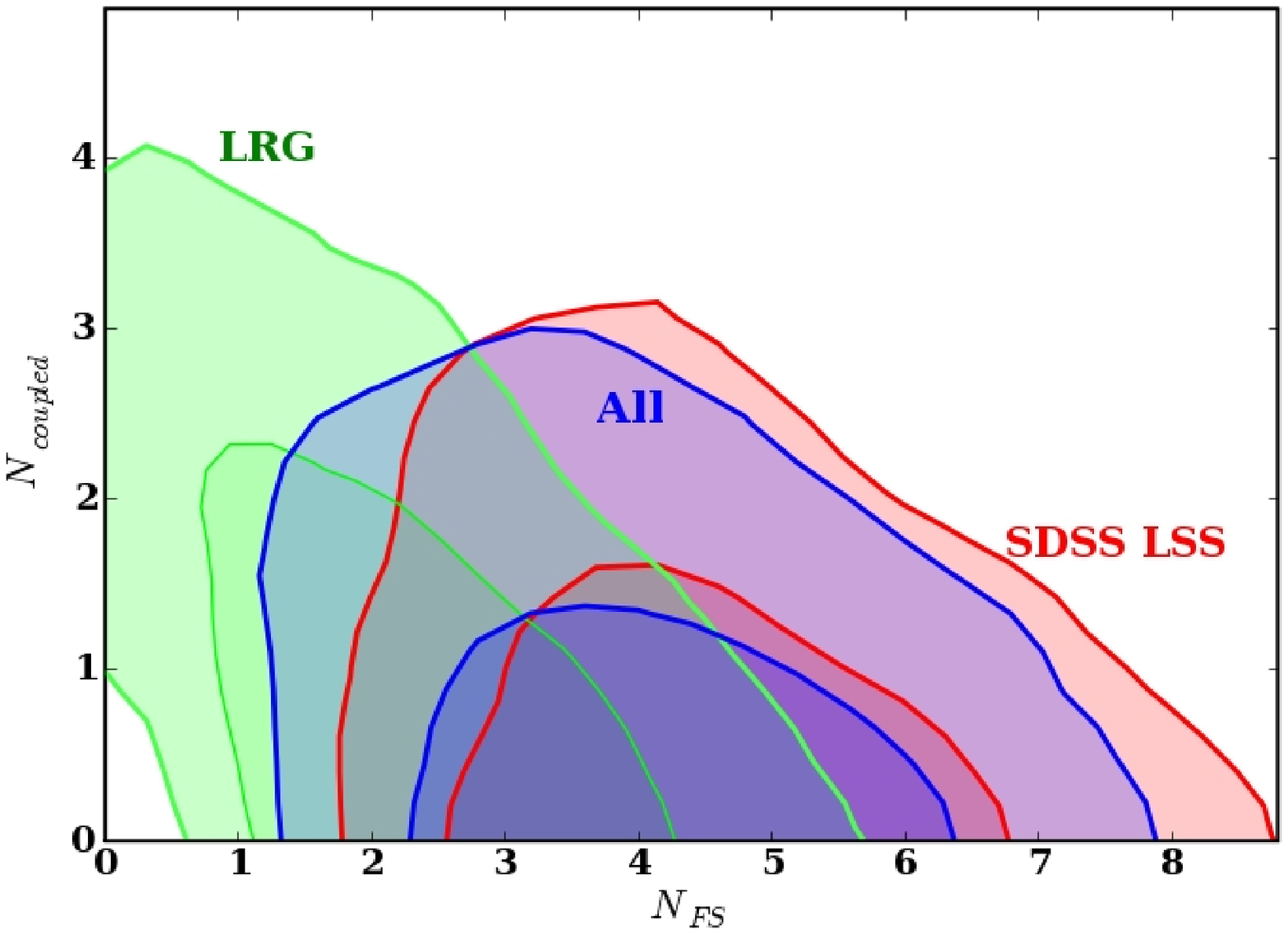}
  \includegraphics[width=0.49\textwidth]{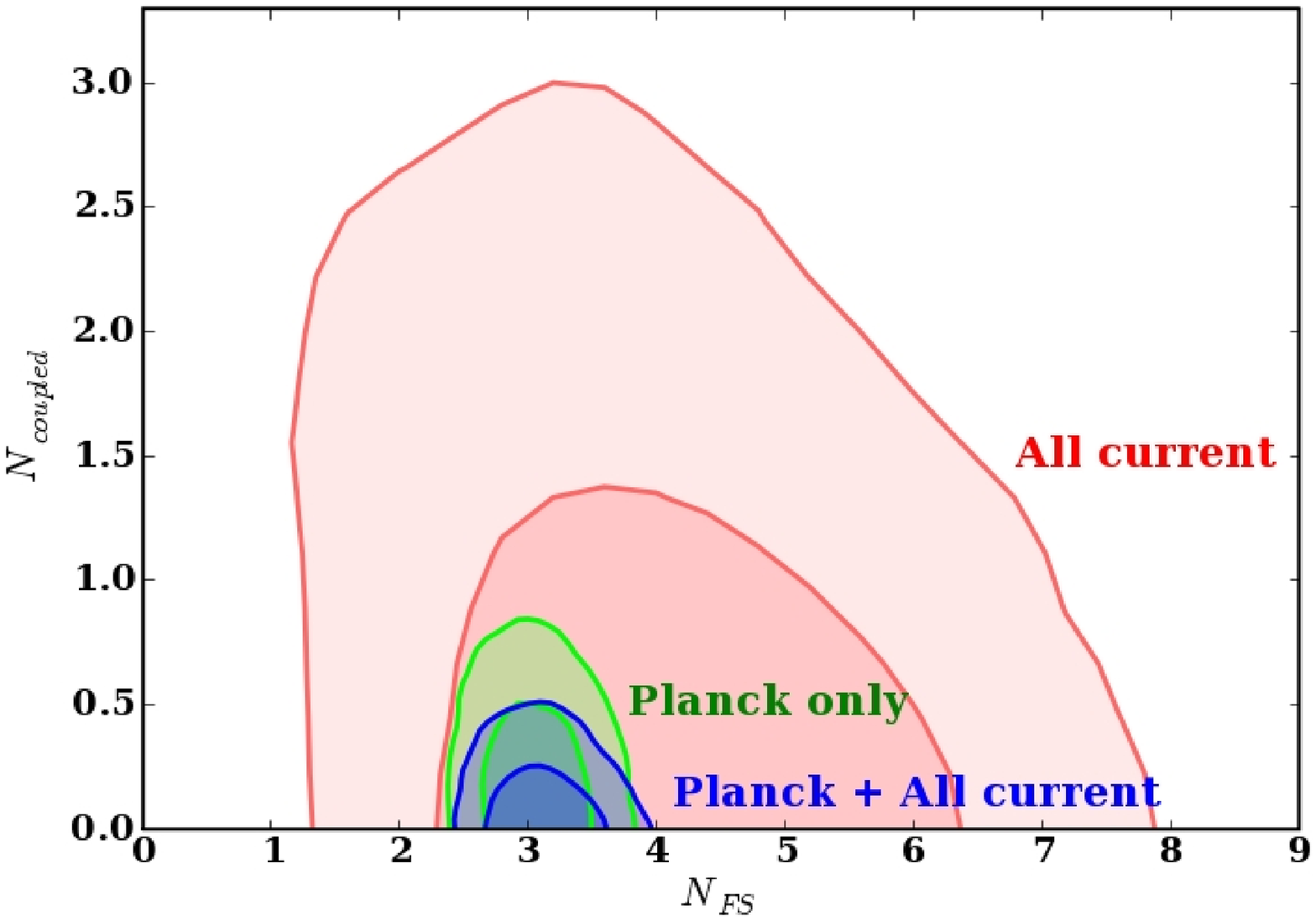}
  \caption{Summary of the bounds on the numbers of free-streaming and
    coupled neutrinos from the current data ({\it left}) and the
    simulated Planck data ({\it right}). For the case of the current
    data, the Large Red Galaxy (LRG) dataset from SDSS yields a region
    that differs significantly from what is obtained with the Main
    SDSS dataset. The combined fit is also shown.  The sensitivity of
    Planck to the numbers of both freely streaming and coupled
    neutrinos will be a dramatic improvement over that of all
    present-day experiments combined (compare ``Planck only'' and
    ``All current'' on the right). If Planck's data is combined
    with today's data, further improvement on the constraints are
    expected.}
  \label{fig:WMAP_vs_Planck}
\end{figure*}

We have considered in this paper the effect of neutrino interactions
on the CMB, both through removal of neutrino free-streaming and the
thermalization of extra relativistic degrees of freedom. The
constraints from the current data are shown in
Figs.~\ref{fig:WMAP_EBCS}, \ref{fig:WMAP_LRG}
and \ref{fig:WMAP3_allbuthLya_scan}; those expected to be obtained with
Planck are depicted in Figs.~\ref{fig:Planck_scanNu_coupled_2d} and
\ref{fig:Planck_all_scanNu_coupled_2d}.  These constraints are
summarized in Fig.~\ref{fig:WMAP_vs_Planck}.

The current data disfavors 3 non-free-streaming neutrinos at the
3.5$\sigma$ level, however, this exclusion is not without some
controversy. The exclusion comes from the SDSS Main data set (and also
from the Lyman-$\alpha$ forest data), which favors five free-streaming
neutrinos. We have seen that the LRG SDSS dataset instead favors close
to three free-streaming neutrinos, in agreement with the standard
model expectations, but has little sensitivity to coupled neutrinos.

We have shown that  Planck will be able to resolve the present controversy.
Planck alone will be capable of excluding a {\em single} interacting
neutrino at the 4.2$\sigma$ level.  Thus we expect sensitivity to models of
neutrino mass and neutrino dark energy which is currently unavailable.
Models of Mass-Varying Neutrinos, for example, in many regions of
parameter space, predict only a single strongly coupled neutrino.
Likewise, models of neutrino mass may have a flavor structure such
that only one or two of the neutrinos is strongly enough coupled to
the scalar to remove neutrino free-streaming at the CMB epoch. For the
MaVaNs models, the reach of Planck will be considerable, probing the
scalar masses down to $m_\phi\sim10^{-8}$ eV, thus covering the
parameter values favored by naturalness.

In the case of extra thermalized neutrinos or scalars, the current constraints depend on which data sets are chosen in the analysis, and in any case the errors encompass multiple neutrinos; as a result, the current data are not conclusive on the total radiation density at the CMB epoch.  By contrast, Planck alone will determine the radiation content to $\Delta N_\nu = ^{+0.5}_{-0.3}$.

The nature of the neutrino mass generation mechanism and beyond-the-standard-model neutrino interactions is currently unknown.   The coupling between a neutrino and a light
scalar in a Majoron interaction term of the form $g \phi \nu \nu$ can
be as small as $g \sim 10^{-13}$ (depending on the mass of the scalar)
while still being large enough to remove neutrino free-streaming and
make an imprint on the CMB.  The constraints (and potential for future detection)  considered in this paper should be a guide to model builders when
considering potential signals from future CMB experiments.  It has been challenging to test neutrino properties in terrestrial experiments, and the nature
of the mechanism which generates neutrino mass remains unknown; future
CMB experiments may have the tools to unmask its origin.

\acknowledgments{ It is our pleasure to thank M.~Perelstein for very
  helpful discussions and comments on the draft. We are also grateful
  to L.~Hall, A.~Lewis, A.~Nelson and A.~Smirnov for valuable
  suggestions and discussions.  This work is supported in part by the
  U.S.  Department of Energy under grant no. DE-FG02-95ER40896.}

\appendix*
\section{On Resonant Production}

In this section we perform a more rigorous derivation of the results of Sect.~VI.

Consider a ``test'' neutrino traveling through a neutrino gas. We
need to know how far it travels before appreciably changing the
direction of its momentum, {\it i.e.}, given the initial momentum
of order $T$, when the particle's momentum in the transverse
direction becomes order $T$. The rate for this process needs to be
compared to the expansion rate of the universe, given by $\sim
T^2/M_{pl}$ in the era of radiation domination.

Let us consider the scattering of two neutrinos,
$\nu\nu\rightarrow\nu\nu$, and start with the $t$-channel exchange of
$\phi$. In this case, for light $\phi$ ($m_\phi\ll T$) the
cross-section is strongly forward peaked (the well-known property of
Rutherford scattering). The situation is well-known in plasma physics.
The build-up of transverse momentum happens mostly as a result of many
small-angle scattering events, rather than a single large-angle event.
The multiple scattering events result in a random walk process for the
transverse momentum $p_T$, so that the rate of change of $p_T^2$ is
given by
\begin{eqnarray}
  \frac{d p_T^2}{dt} \sim \int d(\cos\theta)
  \frac{d\sigma}{d\cos\theta} n v_{rel} p_\theta^2.
\end{eqnarray}
Taking $v_{rel}\sim1$, $p_\theta^2\sim T^2 \theta^2$, $n\sim T^3$ and
$d\sigma/d\cos\theta \sim g^4/(16\pi) T^2(T^2\theta^2+m_\phi^2)^{-2}$,
we get
\begin{eqnarray}
  \frac{d p_T^2}{dt} \sim \frac{g^4}{16\pi}T^3 \ln(T/m_\phi).
\end{eqnarray}
The small-angle behavior of the cross section is regulated by the mass
of the scalar field.
In plasma physics, the logarithm similar to that in the above equation
(``the Coulomb logarithm'') is regulated by plasma screening effects.

The rate for the process needs to be compared to $T^2 \times
T^2/M_{pl}$ (``momentum exchange of order $T$ by the time when the
temperature equals $T$''). We get that neutrinos are streaming
freely at recombination when
\begin{eqnarray}
 \label{eq:offresbound}
  g \lesssim (T_{\rm rec}/M_{pl})^{1/4} \times
 (16\pi/\ln(T_{\rm rec}/m_\phi))^{1/4}.
\end{eqnarray}
The 1/4 power makes the coefficient (the second term) of order one for
a wide range of $m_\phi$, hence for the purpose of this
estimate we can drop it. The order of magnitude is set by the ratio
$(T_{\rm rec}/M_{pl})$. For the temperature $O(eV)$ one thus finds the
neutrinos are not coupled by the $t$-channel scalar exchange if
\begin{eqnarray}
  g \lesssim 10^{-7}.
\end{eqnarray}
The 1/4 power also means that when the $t$ channel process is
dominant it is reasonable to treat neutrinos as tightly coupled on
all scales, not just the sound horizon at the CMB decoupling.
Indeed, even considering scales 100 times smaller than the sound
horizon only changes the bound on $g$ by about a factor of 3.

If $m_\phi>T$, the situation is different. The momentum exchange
occurs via large-angle scattering and the cross section is set by
$T^2/m_\phi^4$, so that
\begin{eqnarray}
  g \gtrsim (T_{\rm rec}/M_{pl})^{1/4}
 (m_\phi/T_{\rm rec}).
\end{eqnarray}
The interaction decouples as the temperature drops, a characteristic
feature of nonrenormalizable interactions. Indeed, at temperature
$T\ll m_\phi$ the $\phi$ field can be integrated out, resulting in an
effective 4-fermion vertex.

We now consider $\nu\nu\rightarrow\nu\nu$ scattering via the
$s$-channel process. In this channel, an important possibility is the
scattering on-resonance. When it is open, it can increase the
sensitivity to $g$ by orders of magnitude, as we will see next.

For $m_\phi<T$, the resonance condition is satisfied when the relative
angle $\theta_{12}$ between the two incoming neutrinos, 1 and 2, is
small, so that $s=E_1 E_2 (1-\cos\theta_{12})=m_\phi^2$, or
$\theta_{12}\sim m_\phi/T$. The relative velocity $v_{rel}$ is given
by $\theta_{12}$. The cross section in the center-of-mass frame is
given by the relativistic Breit-Wigner form,
\begin{eqnarray}
 \sigma_{\rm CM} \sim \frac{\pi}{m_\phi^2}
 \frac{s \Gamma_\phi^2}{(s-m_\phi^2)^2+s^2 \Gamma_\phi^2/m_\phi^2},
\end{eqnarray}
where $\Gamma$ is the width of the resonance, given by the rate of
decay of $\phi$ in its rest frame, $\Gamma_\phi=g^2 m_\phi/(16\pi) $.
The cross section has a peak value $\sim\pi/m_\phi^2$. In the frame of
the universe, the cross section is reduced by a factor of
$m_\phi/T=\theta_{12}$ (since it is boosted in the direction
perpendicular to the direction of the colliding momenta in the
center-of-mass frame).

Collecting all the factors, we obtain
\begin{eqnarray}
  \frac{d p_T^2}{dt} \sim m_\phi^{-2}\theta_{12}
  n_{res} \theta_{12} T^2\theta_{12}^2,
\end{eqnarray}
where $n_{res}$ is the number density of neutrinos on which the test
neutrino scatters resonantly. It can be estimated from the phase space
as $\sim T m_\phi \Gamma_\phi=g^2 T m_\phi^2/(16\pi)$. Hence,
\begin{eqnarray}
 \label{eq:resonantrecoupling}
  \frac{d p_T^2}{dt} \sim \frac{g^2}{16\pi} \frac{m_\phi^4}{T}.
\end{eqnarray}
The result in Eq.~(\ref{eq:resonantrecoupling}) has a simple physical
interpretation \cite{Chacko:2003dt}, as a resonant production of the
$\phi$ particles, $\nu\nu\rightarrow\phi$, completely analogous to the
$Z$-boson production at LEP. Indeed, the rate can be most easily
computed in the opposite direction: the decay of $\phi$ boosted
into the frame of the universe is $g^2 m_\phi^2/T/(16\pi)$. To get the
momentum exchange rate, one recalls that for each
scattering event, one has $p_T \sim m_\phi$.

The crucial feature of the resonant process is the
dependence on $g^2$, rather than $g^4$, since as we saw the reference
values of $g$ are in the $10^{-7}$ range. Instead of
Eq.~(\ref{eq:offresbound}) one has
\begin{eqnarray}
 \label{eq:onresbound}
  g &\lesssim& (4\pi^{1/2})(T_{\rm rec}/M_{pl})^{1/2} \times
  (T_{\rm rec}/m_\phi)^{2} \nonumber\\
  &\lesssim& 10^{-13} (T_{\rm rec}/m_\phi)^{2} .
\end{eqnarray}
For $m_\phi\gtrsim10^{-3}$ eV this constraint is stronger than that in
Eq.~(\ref{eq:offresbound}). Otherwise, the integral over the
nonresonant part of the phase space dominates, giving a bound similar
to Eq.~(\ref{eq:offresbound}).

So far, our discussion has been in complete agreement with
Ref.~\cite{Chacko:2003dt}. We now point out some differences and make
additional observations.

As noted in \cite{Chacko:2003dt}, in the models where the light
scalar $\phi$ is the Goldstone boson there is a possibility to
produce $\phi$ via the $s$-channel diagram with the heavy scalar
$\rho$ (the radial mode) in the intermediate state,
$\nu\bar\nu\rightarrow\rho\rightarrow\phi\phi$. Just like the
process $\nu\nu\rightarrow\phi\rightarrow\nu\nu$ considered above,
this process is most efficient at producing the $phi$ particles on
resonance, {\it i.e.}, at the point in the history of the universe
when the temperature equals $\sim m_\rho$. Unlike the result given in
Eqs.~(28) and (29) of \cite{Chacko:2003dt} we estimate the rate of
this process treating it as $\nu\nu\rightarrow\rho$
 \begin{eqnarray}
  \Gamma \sim \frac{g^2}{32\pi} m_\rho.
\end{eqnarray}
Comparing it with $T^2/M_{pl}=m_\rho^2/M_{pl}$ we find that unless
 \begin{eqnarray}
  g &\lesssim& 4(2\pi)^{1/2} (m_\rho/M_{pl})^{1/2}\nonumber\\
   &\lesssim& 10^{-13} (m_\rho/T_{\rm rec})^{1/2}
\end{eqnarray}
the Goldstone degree of freedom will be populated. If this occurs
before the neutrinos decouple ({\it i.e.}, if $m_\rho\gg 1$ MeV), the
number of effective neutrino degrees of freedom at BBN will be
$3+4/7N_G$ where $N_G$ is the number of Goldstone fields. For a single
Goldstone this may be still acceptable, but more than one would cause
tension with the BBN bounds on $N_{\nu}$ as well as with the projected
bounds of Planck, as we have shown here.

\bibliography{neutrinoCMB0419}

\end{document}